\documentclass[12pt]{article}
\usepackage{epsfig}
\newcommand{\be}{\begin{equation}}
\newcommand{\ee}{\end{equation}}
\newcommand{\bea}{\begin{eqnarray}}
\newcommand{\eea}{\end{eqnarray}}
\newcommand{\nn}{\nonumber}
\newcommand{\norsl}{\normalsize\sl}
\newcommand{\norsc}{\normalsize\sc}

\def \ksl {k \kern-.45em{/}}
\def \lsl {l \kern-.45em{/}}
\def \psl {p \kern-.45em{/}}
\def \Deltasl {\Delta \kern-.65em{/}}
\def \ssl {s \kern-.45em{/}}
\makeatletter
         
          \@addtoreset{equation}{section}
\makeatother

\begin{document}

\begin{titlepage}

\title{Target Mass Effects in Polarized \\
Virtual Photon Structure Functions
}
\author{
\norsc Hideshi BABA\thanks{e-mail address: baba@phys.h.kyoto-u.ac.jp} ,\
     Ken SASAKI\thanks{e-mail address: sasaki@phys.ynu.ac.jp}~ and
           Tsuneo UEMATSU\thanks{e-mail address: 
uematsu@phys.h.kyoto-u.ac.jp} \\
\norsl Graduate School of Human and Environmental Studies,
Kyoto University \\
\norsl Yoshida, Kyoto 606-8501, JAPAN\\
\norsl  Dept. of Physics,  Faculty of Engineering, Yokohama National
University \\
\norsl  Yokohama 240-8501, JAPAN \\
\norsl  Dept. of Physics, Graduate School of Science, Kyoto University \\
\norsl  Yoshida, Kyoto 606-8501, JAPAN \\
}

\date{}
\maketitle

\begin{abstract}
{\normalsize
We study target mass effects in the polarized virtual photon structure 
functions $g_1^\gamma (x,Q^2,P^2)$, $g_2^\gamma (x,Q^2,P^2)$
in the kinematic region $\Lambda^2\ll P^2 \ll Q^2$, where 
$-Q^2 (-P^2)$ is the mass squared of the probe (target) photon.
We obtain the expressions for $g_1^\gamma (x,Q^2,P^2)$ and
$g_2^\gamma (x,Q^2,P^2)$ in closed form by inverting the
Nachtmann moments for the twist-2 and twist-3 operators. Numerical
analysis shows that  target mass effects 
appear at large $x$ and become sizable near $x_{\rm max}(=1/(1+\frac{P^2}{Q^2}))$, the maximal value of  $x$, 
as the ratio  $P^2/Q^2$ increases.
Target mass effects for the sum rules 
of $g_1^\gamma$ and $g_2^\gamma$ are also discussed.
}
\end{abstract}

\begin{picture}(5,2)(-290,-450)
\put(2.3,20){YNU-HEPTh-03-102}
\put(2.3,5){KUNS-1845}
\put(2.3,-10){July 2003}
\end{picture}

\thispagestyle{empty}
\end{titlepage}
\setcounter{page}{1}
\baselineskip 18pt
\section{Introduction}
\smallskip

The investigation of the photon structure is an active field of 
research both theoretically and experimentally~\cite{Kraw,Nisi,Klas,Schi}. 
In recent years, there has been growing interest in the study of the spin 
structure of photon. In particular, the first moment of the polarized 
photon structure function $g_1^\gamma$ has attracted much attention 
in connection with its relevance to the QED and QCD axial
anomaly~\cite{ET,BASS,NSV,FS,BBS}. The polarized photon structure functions 
may be extracted from resolved photon processes in the polarized version of 
the $ep$ collider HERA.  More directly, they can be measured from two-photon
processes in the polarized $e^+e^-$ collider experiments (Fig. 1), where $-Q^2
(-P^2)$ is the  mass squared of the probe (target) photon.

For a real photon ($P^2=0$) target, there exists only one 
spin-dependent structure function $g_1^\gamma(x, Q^2)$. The QCD analysis 
for $g_1^\gamma$ was performed in the leading order (LO) \cite{KS} and 
in the next-to-leading order (NLO) \cite{SV, GRS}. In the case of a virtual 
photon target ($P^2\neq 0$) there appear two spin-dependent structure functions,  
$g_1^\gamma(x, Q^2, P^2)$ and $g_2^\gamma(x, Q^2, P^2)$. The former has been 
investigated up to the NLO in QCD by the present authors in \cite{SU}, and also in
the second paper of \cite{GRS}. In fact, we have analyzed in \cite{SU} the
structure function $g_1^\gamma(x, Q^2, P^2)$ in the kinematical region
$\Lambda^2\ll P^2 \ll Q^2$, where $\Lambda$ is the QCD scale  parameter. The
advantage to study  virtual photon target in that kinematical region is that 
we can calculate structure functions entirely up to the NLO by the
perturbative method~\cite{UW}, which is contrasted with the case of the real
photon  target where in the NLO there exist nonperturbative pieces.
As for the structure function $g_2^\gamma(x, Q^2, P^2)$, the analysis has not made
much progress owing to the  difficulty arising from  the relevant twist-3
operators.  So far only the LO  QCD corrections to the flavor nonsinglet part of
$g_2^\gamma$ have been calculated  in the large  $N_c$ limit \cite{BSU}.

If the target is real photon ($P^2=0$),  there is no need to consider target
mass corrections. But when the target becomes off-shell, for
example, $P^2 \geq M^2$, where $M$ is the nucleon mass, and for relatively 
low values of $Q^2$, contributions suppressed by powers of 
$\frac{P^2}{Q^2}$ may become important. Then we need to
take into account these target mass contributions just like the case of 
the nucleon structure functions. The consideration of target mass effects 
(TME) is important by another reason. 
For the virtual photon target, the maximal value of the Bjorken variable $x$ 
is not 1 but 
\be
x_{\rm max}=\frac{1}{1+\frac{P^2}{Q^2}}~, \label{xmax}
\ee
due to the constraint $(p+q)^2 \ge 0$, which is contrasted with  
the nucleon case where $ x_{\rm max}= 1$.
The structure functions should 
vanish at $x=x_{\rm max}$. However, the NLO QCD result \cite{SU} for   
$g_1^\gamma(x, Q^2, P^2)$ shows that the predicted graph does not vanish but
remains finite  at $x=x_{\rm max}$. This flaw is coming from the fact that
 TME have not been taken into account  in the analysis.  
The target mass corrections have been studied in the past for the cases of unpolarized
~\cite{NACHT,GP} and polarized~\cite{WAND,MU,KU,PR,BK} nucleon structure functions. 

In this paper we investigate   
TME in  the polarized virtual photon structure functions 
$g_1^\gamma (x,Q^2,P^2)$ and $g_2^\gamma (x,Q^2,P^2)$. In the analysis of
$g_1^\gamma$ in \cite{SU}, the formalism of the operator product expansion (OPE) 
supplemented by  the  renormalization group method was used. The photon matrix
elements of the relevant traceless operators in the OPE are expressed by 
traceless tensors.  These tensors contain many trace terms 
so that they satisfy the tracelessness conditions. The basic idea for computing 
the target mass corrections exactly is to take account of trace terms in the 
traceless tensors properly. There are two methods used so far for collecting 
all those trace terms. One, which was introduced by Nachtmann~\cite{NACHT}, 
is to make use of Gegenbauer polynomials to express the contractions 
between $q_{\mu_1}\cdots q_{\mu_{n-1}}$ and the traceless tensors~\cite{NACHT, WAND,
MU, KU}. This method  leads to the Nachtmann moments for the twist-2 and twist-3
operators  with definite spin. 
The other, first used by Georgi and Politzer~\cite{GP}, is to write  
traceless tensors explicitly and then to collect trace terms and sum them up. 
Through the latter approach, the moments of structure functions are expressed 
as functions of the reduced operator matrix elements
and coefficient functions with different spins. Actually both methods give equivalent
results.  In this paper we apply the former method to study  target mass
corrections to the structure functions  $g_1^\gamma$ and $g_2^\gamma$. 

In the next section we discuss the framework for analyzing 
the TME based on the OPE and derive the Nachtmann moments for the twist-2 and twist-3
operators  with definite spin using
the orthogonality relations of Gegenbauer polynomials. In section 3, by
inverting the Nachtmann moments, we obtain
the explicit expressions for the polarized photon structure functions 
$g_1^\gamma (x,Q^2,P^2)$ and $g_2^\gamma (x,Q^2,P^2)$ with TME included.
In section 4 we perform the numerical analysis and show that target mass
corrections become sizable near $x_{\rm max}$. 
The final section is devoted to the conclusion.

\section{Nachtmann Moments}
\smallskip
Let us consider the virtual photon-photon forward scattering for
 $\gamma(q)+\gamma(p)\rightarrow \gamma(q)+\gamma(p)$ 
illustrated in Fig.2,
\be
T_{\mu\nu\rho\tau}(p,q)=i\int d^4 x d^4 y d^4 z e^{iq\cdot x}e^{ip\cdot (y-z)}
\langle 0|T(J_\mu(x) J_\nu(0) J_\rho(y) J_\tau(z))|0\rangle.
\ee
where $J$ is the electromagnetic current, and $q$ and $p$ are the
four-momenta of two photons.  
Its absorptive part is related to the structure tensor 
$W_{\mu\nu\rho\tau}(p,q)$ for the target photon with mass squared 
$p^2=-P^2$ probed by the photon with $q^2=-Q^2$:
\be
W_{\mu\nu\rho\tau}(p,q)=\frac{1}{\pi}{\rm Im}T_{\mu\nu\rho\tau}(p,q)~.
\ee
The antisymmetric part $W_{\mu\nu\rho\tau}^{A}$ under the interchange 
$\mu\leftrightarrow\nu$ and $\rho\leftrightarrow\tau$,  can be decomposed as
\bea
W_{\mu\nu\rho\tau}^{A}&=&
\epsilon_{\mu\nu\lambda\sigma}q^\lambda
{\epsilon_{\rho\tau}}^{\sigma\beta}p_\beta \frac{1}{p\cdot
q}g_1^\gamma\nonumber\\
&+&
\epsilon_{\mu\nu\lambda\sigma}q^\lambda
(p\cdot q\ {\epsilon_{\rho\tau}}^{\sigma\beta}p_\beta-\epsilon_{\rho\tau
\alpha\beta}p^\beta p^\sigma q^\alpha )\frac{1}{(p\cdot q)^2}g_2^\gamma~,
\eea
which gives two spin-dependent structure functions, $g_1^\gamma(x,Q^2,P^2)$ 
and $g_2^\gamma(x,Q^2,P^2)$. When the target is real photon ($P^2= 0$), $g_2^\gamma$
is identically  zero, and there exists only one spin structure function,
$g_1^\gamma(x,Q^2)$.  On the other hand, for the off-shell or virtual photon ($P^2
\neq 0$) target,  we have two 
spin-dependent structure functions $g_1^\gamma$ and $g_2^\gamma$.

For the analysis of spin structure functions, we 
apply the OPE for the product of two
electromagnetic currents.  We obtain for the $\mu$-$\nu$ antisymmetric part
\bea
&&i\int d^4x e^{iq\cdot x}T(J_\mu(x)J_\nu(0))^A
=-i\epsilon_{\mu\nu\lambda\sigma}q^\lambda
\sum_{n=1,3,\cdots}\left(\frac{2}{Q^2}\right)^n
q_{\mu_1}\cdots q_{\mu_{n-1}}\nonumber\\
&&\hspace{4cm}\times
\left\{
\sum_i E_{(2)i}^n R_{(2)i}^{\sigma\mu_1\cdots\mu_{n-1}}
+\sum_i E_{(3)i}^n R_{(3)i}^{\sigma\mu_1\cdots\mu_{n-1}}
\right\}~,\label{CurrentProdFourier}
\eea
where $R^n_{(2)i}$ and $R^n_{(3)i}$ are the twist-2 and twist-3 operators,
respectively, and  are both traceless, and $E_{(2)i}^n$ and $E_{(3)i}^n$ are 
corresponding coefficient functions.
The twist-2 operators $R^n_{(2)i}$ have totally symmetric Lorentz indices
$\sigma\mu_1\cdots\mu_{n-1}$, while the indices of twist-3 operators
$R^n_{(3)i}$  are totally symmetric among $\mu_1\cdots\mu_{n-1}$ but
antisymmetric under $\sigma \leftrightarrow \mu_i$. 

In the case of photon target we evaluate ``{\it matrix elements}" of the traceless 
operators $R^n_{(2)i}$ and $R^n_{(3)i}$  sandwiched by 
two photon states with momentum $p$,  which are written in the following forms:
\bea
\langle 0\vert T(A_{\rho}(-p)R_{(2)i}^{\sigma\mu_1\cdots\mu_{n-1}}A_{\tau}(p))
\vert 0\rangle_{\rm Amp}&=&-ia_{(2)i}^{\gamma,n}
M_{(2)\rho\tau}^{\sigma\mu_1\cdots\mu_{n-1}}~, \label{matTwist2}\\
\langle 0\vert T(A_{\rho}(-p)R_{(3)i}^{\sigma\mu_1\cdots\mu_{n-1}}A_{\tau}(p))
\vert 0\rangle_{\rm Amp}&=&-ia_{(3)i}^{\gamma,n}
M_{(3)\rho\tau}^{[\sigma,\{\mu_1]\cdots\mu_{n-1}\}}~,\label{matTwist3}
\eea
where the subscript \lq Amp\rq\ stands for the amputation of  external
photon lines, $a_{(2)i}^{\gamma,n}$ and $a_{(3)i}^{\gamma,n}$  are reduced 
photon matrix elements. The tensors $M_{(2)\rho\tau}^{\sigma\mu_1\cdots\mu_{n-1}}$ and
$M_{(3)\rho\tau}^{[\sigma,\{\mu_1]\cdots\mu_{n-1}\}}$ are given by
\bea
M_{(2)\rho\tau}^{\sigma\mu_1\cdots\mu_{n-1}}
 &\equiv& \frac{1}{n}\Bigl[
{\epsilon_{\rho\tau\alpha}}^\sigma p^{\mu_1}\cdots
p^{\mu_{n-1}} +\sum^{n-1}_{j=1}p^\sigma
p^{\mu_1}\cdots {\epsilon_{\rho\tau\alpha}}^{\mu_j}\cdots p^{\mu_{n-1}} \Bigr]p^\alpha\nn\\
&&\hspace{6cm}-({\rm trace\ terms})~,  \label{formTwist2}\\
M_{(3)\rho\tau}^{[\sigma,\{\mu_1]\cdots\mu_{n-1}\}}
  &\equiv&\biggl[ \frac{n-1}{n}
{\epsilon_{\rho\tau\alpha}}^{\sigma}p^{\mu_1}\cdots
p^{\mu_{n-1}}-\frac{1}{n}\sum^{n-1}_{j=1}p^{\sigma}
p^{\mu_1}\cdots {\epsilon_{\rho\tau\alpha}}^{\mu_j}\cdots p^{\mu_{n-1}}\biggr]p^\alpha \nn \\
&& \hspace{6cm}-({\rm trace\ terms}) ~, \label{formTwist3}
\eea
and satisfy the traceless conditions, 
\be
  g_{\sigma\mu_i}M_{(k)\rho\tau}^{\sigma\mu_1\cdots\mu_{n-1}}=0, \qquad
  g_{\mu_i\mu_j}M_{(k)\rho\tau}^{\sigma\mu_1\cdots\mu_{n-1}}=0 \qquad (k=2,3)~.
\ee
Taking the \lq\lq {\it matrix elements}\rq\rq\ of
(\ref{CurrentProdFourier}) with the virtual photon states,
we obtain for the deep-inelastic photon-photon forward scattering amplitude
\bea
T^A_{\mu\nu\rho\tau}&=&
i\int d^4x e^{iq\cdot x}
\langle 0|T(A_\rho(-p)(J_\mu(x)J_\nu(0))^AA_\tau(p))|0\rangle_{\rm Amp}
\nonumber\\
&=&-\epsilon_{\mu\nu\lambda\sigma}q^\lambda\sum_{n=1,3,\cdots}
\left(\frac{2}{Q^2}\right)^n q_{\mu_1}\cdots q_{\mu_{n-1}}
\nonumber\\
&&
\hspace{1.3cm}\times\left\{\sum_i a_{(2)i}^{\gamma,n}E_{(2)i}^n
M_{(2)\rho\tau}^{\sigma\mu_1\cdots\mu_{n-1}}+ a_{(3)i}^{\gamma,n}E_{(3)i}^n
M_{(3)\rho\tau}^{[\sigma,\{\mu_1]\cdots\mu_{n-1}\}} \right\}.\label{Amplitude}
\eea

The basic idea for treating  target mass corrections exactly is 
to take account of trace terms in the traceless tensors properly.
We evaluate the contraction between $q_{\mu_1}\cdots q_{\mu_{n-1}}$ and the traceless 
tensors without neglecting any of the trace terms in
Eqs.(\ref{formTwist2}) and (\ref{formTwist3}).  The results are expressed in terms of
Gegenbauer polynomials
\cite{NACHT, WAND, MU}.
Denoting $M_{(2)\rho\tau}^{\sigma\mu_1\cdots\mu_{n-1}}
\equiv{{\widetilde M}_{(2)\beta}}^{\sigma\mu_1\cdots\mu_{n-1}}
{\epsilon_{\rho\tau\alpha}}^\beta p^\alpha$ and
$M_{(3)\rho\tau}^{[\sigma,\{\mu_1]\cdots\mu_{n-1}\}}
\equiv{{\widetilde M}_{(3)\beta}}^{\sigma\mu_1\cdots\mu_{n-1}}
{\epsilon_{\rho\tau\alpha}}^\beta p^\alpha$, we find for the twist-2 part,
\bea
q_{\mu_1}\cdots q_{\mu_{n-1}} 
{{\widetilde M}_{(2)\beta}}^{\sigma\mu_1\cdots\mu_{n-1}}
 &=&\frac{1}{n^2}\left[
{\delta_\beta}^\sigma a^{n-1} C_{n-1}^{(2)}(\eta)+q_\beta p^\sigma a^{n-2}
~ 2C_{n-2}^{(3)}(\eta)\right] \nn\\
&&+({\rm terms\ with}\ p_\beta\  {\rm or}\ q^\sigma), \label{cont-tw2}
\eea
and for the twist-3 part
\bea
q_{\mu_1}\cdots q_{\mu_{n-1}} 
{{\widetilde M}_{(3)\beta}}^{\sigma\mu_1\cdots\mu_{n-1}}
&=&{\delta_\beta}^\sigma \frac{a^{n-1}}{n^2}
\left[(n-1)C_{n-1}^{(2)}(\eta)-(n+1)C_{n-3}^{(2)}(\eta)\right]\nonumber\\
&&-q_\beta p^\sigma \frac{2a^{n-2}}{n^2}\left[
C_{n-2}^{(3)}(\eta)+C_{n-4}^{(3)}(\eta)\right]\nn\\
&& +({\rm terms\ with}\ p_\beta\  {\rm or}\ q^\sigma),
\label{cont-tw3}
\eea
where $a=-\frac{1}{2}PQ$, $\eta=-p\cdot q/PQ$ and $C_n^{(\nu)}(\eta)$'s are 
Gegenbauer polynomials.  In fact in the above two equations there appear
terms with $p_\beta$ or $q^\sigma$. These 
terms  give null results  when they are multiplied by 
${\epsilon_{\rho\tau\alpha}}^\beta p^\alpha$ 
and  $\epsilon_{\mu\nu\lambda\sigma}q^\lambda$. 
(See Appendix for the derivation of Eqs.(\ref{cont-tw2}-\ref{cont-tw3})).

We decompose the amplitude $T^A_{\mu\nu\rho\tau}$ as 
\be
T^A_{\mu\nu\rho\tau}=
\epsilon_{\mu\nu\lambda\sigma}q^\lambda 
{\epsilon_{\rho\tau}}^{\sigma\beta}p_\beta \frac{1}{p\cdot q}(v_1^\gamma+v_2^\gamma)
-\epsilon_{\mu\nu\lambda\sigma}q^\lambda p^\sigma
\epsilon_{\rho\tau\alpha\beta}q^\alpha p^\beta \frac{1}{(p\cdot q)^2}v_2^\gamma~,
\label{Decomposition}
\ee
then, using the above results on the contractions we find
\bea
v_1^\gamma+v_2^\gamma&=&\sum_{n=1,3,\cdots}\sum_i a_{(2)i}^{\gamma,n}E_{(2)i}^n
\left(-\frac{P}{Q}\right)^n \frac{1}{n^2}2\eta C_{n-1}^{(2)}(\eta) \nn\\
&&-\sum_{n=3,5\cdots}\sum_i  a_{(3)i}^{\gamma,n}E_{(3)i}^n
\left(-\frac{P}{Q}\right)^n \frac{1}{n^2}2\eta 
\left((n+1)C_{n-3}^{(2)}(\eta)-(n-1)C_{n-1}^{(2)}(\eta) \right) \label{v1v2}\nn\\
&&\\
v_2^\gamma&=&-\sum_{n=1,3,\cdots}\sum_i a_{(2)i}^{\gamma,n}E_{(2)i}^n
\left(-\frac{P}{Q}\right)^n \frac{1}{n^2}8\eta^2 C_{n-2}^{(3)}(\eta)\nn\\
&&+\sum_{n=3,5\cdots}\sum_i  a_{(3)i}^{\gamma,n}E_{(3)i}^n
\left(-\frac{P}{Q}\right)^n
\frac{1}{n^2}8\eta^2\left(C_{n-2}^{(3)}(\eta)+C_{n-4}^{(3)}(\eta)\right)~.\label{v2}
\eea

Here we compare the expressions of $v_1^\gamma+v_2^\gamma$ and $v_2^\gamma$ 
with those given in Eqs.(8) and (9) of Ref.\cite{MU}, which are
the invariant amplitudes for polarized deep inelastic lepton-nucleon 
scattering  with  target nucleon mass corrections being taken into account. 
Apart from the reduced matrix elements and  coefficient functions, the expressions
for  both  photon and nucleon targets are exactly the same once the 
replacement of $M$  with $-iP$, or {\it vice versa},  is made. 
This is due to the fact that the factor $[{\epsilon_{\rho\tau\alpha}}^\sigma p^\alpha]
~ ([{\epsilon_{\rho\tau\alpha}}^{\mu_j}p^\alpha] )$ appearing in the 
photon  matrix elements (Eqs.(\ref{formTwist2}) and (\ref{formTwist3})) 
and in the decomposition (Eq.(\ref{Decomposition})) plays the same role as 
nucleon spin $s^\sigma (s^{\mu_j})$, since $p_\sigma{\epsilon_{\rho\tau\alpha}}^\sigma
p^\alpha=0~ (p_{\mu_j}{\epsilon_{\rho\tau\alpha}}^{\mu_j}p^\alpha=0)$. Thus 
the tensor structures of both  polarized photon and polarized nucleon matrix elements
are exactly the same. The only difference between the two is that $p^2=-P^2$ for 
photon target and $p^2=M^2$  for nucleon.

Now we follow the same procedures as were taken by Wandzura~\cite{WAND} and
in Ref.~\cite{MU}  for the polarized nucleon case, and we obtain the
analytic expression of  the Nachtmann moments for the twist-2 and twist-3
operators with definite spin $n$. First we write the dispersion relations for
$v_1^\gamma$ and
$v_2^\gamma$ and denote 
\be
g_{1,2}^\gamma=\frac{1}{\pi}{\rm Im}\,v_{1,2}^\gamma~~.
\ee
Secondly using the orthogonality relations (Eq.(\ref{orthogonal})) and an integration
formula (Eq.(\ref{IntegralGegenbauer})) for  Gegenbauer polynomials
$C_{n}^{(\nu)}(\eta)$, we project out $\sum_i a_{(2)i}^{\gamma,n}E_{(2)i}^n$ and 
$\sum_i a_{(3)i}^{\gamma,n}E_{(3)i}^n$ with definite spin $n$, which still include 
the infinite series in powers of $P^2/Q^2$. Thirdly we sum up those infinite series 
and express them in compact analytic forms \cite{MU}. Then we obtain
\bea
M_2^n&\equiv& \sum_i a_{(2)i}^{\gamma,n}E_{(2) i}^n(Q^2,P^2,g)\nonumber\\
&=&\int_0^{x_{\rm max}}\frac{dx}{x^2}{\xi}^{n+1}\left[
\left\{
\frac{x}{\xi}+\frac{n^2}{(n+2)^2}\frac{P^2 x\xi}{Q^2}
\right\}g_1^\gamma(x,Q^2,P^2)\right.    \label{Nacht2}\\
&&\left. \hspace{5cm}+\frac{4n}{n+2}\frac{P^2x^2}{Q^2}g_2^\gamma(x,Q^2,P^2)
\right] , \quad (n=1,3,\cdots) \nn  \\
M_3^n &\equiv& \sum_i a_{(3)i}^{\gamma,n}E_{(3) i}^n(Q^2,P^2,g)\nonumber\\
&=&\int_0^{x_{\rm max}}\frac{dx}{x^2}{\xi}^{n+1}\left[
\frac{x}{\xi}g_1^\gamma(x,Q^2,P^2)\right.\label{Nacht3}\\
&&\left.\hspace{2.5cm}+\left\{\frac{n}{n-1}\frac{x^2}{\xi^2}+
\frac{n}{n+1}\frac{P^2x^2}{Q^2} \right\}g_2^\gamma(x,Q^2,P^2) \right] 
 , \quad (n=3,5, \cdots) \nn
\eea
where $x=Q^2/(2p\cdot q)$ and $\xi$, the so-called $\xi$-scaling
variable, is given by
\be
\xi=\frac{2x}{1+\sqrt{1-\frac{4P^2x^2}{Q^2}}}~. \label{xi}
\ee
In fact, the above results for the Nachtmann moments are reproduced from the
counterparts in the case of spin-dependent nucleon structure functions,   Eqs. (18)
and (19) of Ref.\cite{MU}, by replacing the target nucleon mass $M$ with $-iP$. 

We see from Eq.(\ref{xmax}) that the maximal value of $x$ is not 1 but 
$1/(1+\frac{P^2}{Q^2})$. Therefore,  the allowed range of $\xi$ is  $0\le \xi\le1$. 
It is important to note that  $\xi(x_{\rm max})=1$  for the virtual photon target. 
In the nucleon case,  however, 
the constraint $(p+q)^2\geq M^2$ gives $x_{\rm max}=1$. Changing $P^2$ in 
Eq.(\ref{xi}) to $-M^2$, we get that
$\xi(x=1)<1$. This leads to a well-known difficulty at $x=1$ in the
analysis  of target mass corrections to nucleon structure functions, both in 
unpolarized and polarized cases. The nucleon structure 
functions should vanish at $x=1$ kinematically, while their expressions, once
target mass effects are taken into account, 
vanish at $\xi=1$ but remain non-zero when $\xi(x=1)<1$.  
The resolution to this problem was argued in Refs.\cite{RGP,PR,BK}
by considering the dynamical higher-twist effects.
On the other hand, 
in the case of virtual photon target, we have no such difficulty. When 
$Q^2,P^2\gg \Lambda^2$, we can  put the constraint
as $(p+q)^2 \geq 0$ and this leads to $x_{\rm max}$ given in Eq. (\ref{xmax}). 
We will see later that the virtual photon structure functions with target mass 
corrections included do vanish at $x_{\rm max}$, since $\xi(x_{\rm max})=1$.

The left-hand side of Eqs.(\ref{Nacht2}-\ref{Nacht3}), $M_2^n$ and $M_3^n$, can be
computed in perturbative QCD up to NLO,  since in the kinematical 
region $Q^2\gg P^2\gg \Lambda^2$ both the reduced photon matrix elements 
$a_{(k)i}^{\gamma,n}~ (k=2,3)$ and coefficient functions $E_{(k) i}^n(Q^2,P^2,g)$ $(k=2,3)$ 
are calculable.  In fact, the perturbative QCD calculation of $M_2^n$ has been done in
LO~\cite{KS} and  in NLO~\cite{SU}, while the QCD analysis of $M_3^n$ has been
carried out in LO for the flavor non-singlet part in the limit of large $N_c$~\cite{BSU}.
Once the moments $M_2^n$ and $M_3^n$ are known, we can derive 
$g_1^\gamma(x,Q^2,P^2)$ and $g_2^\gamma(x,Q^2,P^2)$ as functions of $x$ by
inverting  $M_2^n$ and $M_3^n$, which will be discussed in the
next section.
\bigskip
\section{Inverting the Moments}
\smallskip
First let us rewrite the Nachtmann moments in the variable $\xi$ and we get
\bea
M_2^n&=&\int_0^1d\xi\ \xi^{n-1}\biggl[\left\{1+\frac{n^2}{(n+2)^2}\kappa\xi^2  
\right\} \frac{1-\kappa\xi^2}{1+\kappa\xi^2}~g_1^\gamma(x,Q^2,P^2)
\nn \\
&&\qquad \qquad \qquad+\frac{n}{n+2}4 \kappa\xi^2
\frac{1-\kappa\xi^2}{(1+\kappa\xi^2)^2}~g_2^\gamma(x,Q^2,P^2)
\biggr]\\
&&\nn\\
M_3^n&=&\int_0^1d\xi\ \xi^{n-1}\biggl[
\frac{1-\kappa\xi^2}{1+\kappa\xi^2}~g_1^\gamma(x,Q^2,P^2)
\nn \\
&&\qquad \qquad +\left\{  \frac{n}{n-1}+\frac{n}{n+1} \kappa\xi^2\right\}
\frac{1-\kappa\xi^2}{(1+\kappa\xi^2)^2}~g_2^\gamma(x,Q^2,P^2)
\biggr]
\eea
where we have put $\kappa\equiv \frac{P^2}{Q^2}$. We define
\be
A(\xi)\equiv \frac{1-\kappa\xi^2}{1+\kappa\xi^2}~g_1^\gamma(x,Q^2,P^2)~, \quad 
B(\xi)\equiv \frac{1-\kappa\xi^2}{(1+\kappa\xi^2)^2}~g_2^\gamma(x,Q^2,P^2)~.
\ee 
The boundary conditions for $A(\xi)$ and $B(\xi)$ are
$A(\xi=1)=B(\xi=1)=0$, since $g_1^\gamma(x_{\rm max},Q^2,P^2)=g_2^\gamma(x_{\rm
max},Q^2,P^2)=0$ and  $\xi(x_{\rm max})=1$. 
Now introducing the following four functions,
\bea
{\widetilde A}(\xi)&=&\int^1_\xi \frac{d\xi'}{\xi'}\int^1_{\xi'}\frac{d\xi''}{\xi''}A(\xi'')\\
B_{-1}(\xi)=\int^1_\xi \frac{d\xi'}{\xi'}B(\xi'), \quad
B_{0}(\xi)&=&\int^1_\xi d\xi'B(\xi'), \quad B_{1}(\xi)=\int^1_\xi d\xi'\{\xi' B(\xi')\},\nn\\
\eea
and by partial integration we find that the above two moments are written as
\bea
\frac{M_2^n}{n^2}
&=&\int_0^1d\xi~\xi^{n-1}\Bigl[\Bigl(1+\kappa\xi^2\Bigr){\widetilde A}(\xi)
+2\kappa\Bigl\{ B_{1}(\xi)-\xi^2  B_{-1}(\xi)  \Bigr\}\Bigr],\\
&&\nn\\
\frac{M_3^n}{n^2}
&=&\int_0^1d\xi~\xi^{n-1}\Bigl[{\widetilde A}(\xi)+\frac{1}{\xi}
\Bigl(1-\kappa\xi^2\Bigr)B_0(\xi) -B_{-1}(\xi)+\kappa B_1(\xi)\Bigr]~.
\eea
Inverting the moments we get
\bea
H_a(\xi)&=&\frac{1}{2\pi i}\int_{c-i\infty}^{c+i\infty}
dn \,\xi^{-n}\biggl\{ \frac{M_2^n}{n^2} \biggr\} \nn\\
&=&\Bigl(1+\kappa\xi^2\Bigr){\widetilde A}(\xi)
+2\kappa\Bigl\{ B_{1}(\xi)-\xi^2  B_{-1}(\xi)  \Bigr\},
\label{InversionTwist2}\\ &&\nn\\
H_d(\xi)&=&\frac{1}{2\pi i}\int_{c-i\infty}^{c+i\infty}
dn \,\xi^{-n}\biggl\{ \frac{M_3^n}{n^2} \biggr\} \nn\\
&=&{\widetilde A}(\xi)+\frac{1}{\xi}
\Bigl(1-\kappa\xi^2\Bigr)B_0(\xi) -B_{-1}(\xi)+\kappa B_1(\xi),\label{InversionTwist3}
\eea
where we have adopted the notation used in Ref.\cite{PR} for $H_a(\xi)$ and $H_d(\xi)$.
Further introducing the following functions~\cite{PR},
\be
G_{a,d}(\xi)=-\xi\frac{dH_{a,d}(\xi)}{d\xi}~, \qquad F_{a,d}(\xi)=-\frac{dG_{a,d}(\xi)}{d\xi}~,
\label{GFad}
\ee
we differentiate both sides of Eqs.(\ref{InversionTwist2}-\ref{InversionTwist3}) 
by $\xi$ and get the relations between ${\widetilde A}(\xi)$, $B_{-1}(\xi)$,
$B_{0}(\xi)$, $B_{1}(\xi)$ and $H_{a,d}(\xi)$, $G_{a,d}(\xi)$, $F_{a,d}(\xi)$.
Now replacing  the former functions with the latter,  
we solve for  $g_1^\gamma$ and $g_2^\gamma$ and obtain
\bea
g_1^\gamma(x,Q^2,P^2)&=&4\kappa\xi^2~
\frac{(1+\kappa\xi^2)^3}{(1-\kappa\xi^2)^5}\left\{
1+\frac{2\kappa\xi^2}{(1+\kappa\xi^2)^2}  \right\}H_a(\xi)\nn\\
&&-4\kappa\xi^2~
\frac{(1+\kappa\xi^2)^2}{(1-\kappa\xi^2)^4}\left\{
1+\frac{1}{1+\kappa\xi^2}  \right\}G_a(\xi)
+\xi~\frac{(1+\kappa\xi^2)^2}{(1-\kappa\xi^2)^3}F_a(\xi)\nn\\
&&-8\kappa\xi^2~
\frac{(1+\kappa\xi^2)^3}{(1-\kappa\xi^2)^5}\left\{
1+\frac{2\kappa\xi^2}{(1+\kappa\xi^2)^2}  \right\}H_d(\xi)\nn\\
&&+12\kappa\xi^2~
\frac{(1+\kappa\xi^2)^2}{(1-\kappa\xi^2)^4}G_d(\xi)
-4\kappa\xi^3~\frac{1+\kappa\xi^2}{(1-\kappa\xi^2)^3}F_d(\xi) \label{TargetMass2}\\
&&\nn\\
g_2^\gamma(x,Q^2,P^2)&=&-6\kappa\xi^2~
\frac{(1+\kappa\xi^2)^3}{(1-\kappa\xi^2)^5}H_a(\xi)
+\frac{(1+\kappa\xi^2)^3}{(1-\kappa\xi^2)^4}\left\{
1+\frac{4\kappa\xi^2}{1+\kappa\xi^2}  \right\}G_a(\xi)\nn\\
&&-\xi~\frac{(1+\kappa\xi^2)^2}{(1-\kappa\xi^2)^3}F_a(\xi)\nn\\
&&+12\kappa\xi^2~
\frac{(1+\kappa\xi^2)^3}{(1-\kappa\xi^2)^5}H_d(\xi)
-\frac{(1+\kappa\xi^2)^4}{(1-\kappa\xi^2)^4}
\left\{1+\frac{8\kappa\xi^2}{(1+\kappa\xi^2)^2}  \right\}
G_d(\xi)\nn\\
&&+\xi\frac{(1+\kappa\xi^2)^3}{(1-\kappa\xi^2)^3}F_d(\xi)~. \label{TargetMass3}
\eea

Eqs.(\ref{TargetMass2}), (\ref{TargetMass3}) are the final formulas for the polarized photon
structure functions $g_1^\gamma$ and $g_2^\gamma$ when  target mass  effects are taken into
account. The parameter
$\kappa$ represents  the target mass corrections. Once the reduced photon matrix elements and 
coefficient functions corresponding to the relevant twist-2 and -3 operators, more specifically, 
$\sum_i a_{(2)i}^{\gamma,n}E_{(2) i}^n(Q^2,P^2,g)$ and $\sum_i a_{(3)i}^{\gamma,n}E_{(3)
i}^n(Q^2,P^2,g)$ in Eqs.(\ref{Nacht2}-\ref{Nacht3}), are given, then we can calculate
$H_{a,d}(\xi)$, $G_{a,d}(\xi)$ and  $F_{a,d}(\xi)$ through
Eqs.(\ref{InversionTwist2}-\ref{GFad}), and predict whole structure functions with target
mass corrections. Note that by definition the functions $H_{a,d}(\xi)$,
$G_{a,d}(\xi)$ and  $F_{a,d}(\xi)$ contain the logarithmic QCD corrections 
depending on ${\ln}(Q^2/\Lambda^2)$ and ${\ln}(P^2/\Lambda^2)$. 
When we set $\kappa=0$ in Eqs.(\ref{TargetMass2}-\ref{TargetMass3}) and Eq.(\ref{xi}), 
we obtain
\bea
g_1^\gamma(x,Q^2,P^2)\vert_0&=&x F_a(x)  \label{g1WithoutTME}\\
g_2^\gamma(x,Q^2,P^2)\vert_0&=&G_a(x) -x F_a(x) -G_d(x)+x F_d(x)
\label{g2WithoutTME}
\eea
for the polarized photon structure functions without 
target mass effects, which have been investigated in the literature \cite{SU,GRS,BSU}. 
We have suppressed the logarithmic $Q^2$ and $P^2$ dependence
in the arguments of $G_{a,d}(x)$ and $F_{a,d}(x)$.

Before we proceed to numerical analysis for target mass effects on the 
polarized photon structure functions, let us consider the power series 
expansion of target mass effects.
In the phenomenological analysis of target mass effects on the polarized
nucleon structure functions \cite{PR}, the expansion in powers of $P^2/Q^2$
was carried out and the first order terms were kept to analyze the experimental
data. It would be interesting to see how good the first order approximation
is in the case of virtual photon target.
We take the $x$-moments of the structure functions $g_{1,2}^\gamma(x,Q^2,P^2)$,
\be
g_{1,2}^{\gamma,n}\equiv \int_0^{x_{\rm max}} dx x^{n-1}
g_{1,2}^\gamma(x,Q^2,P^2)=\int_0^1 d\xi
\frac{1-\kappa\xi^2}{(1+\kappa\xi^2)^2}\left(\frac{\xi}{1+\kappa\xi^2}\right)^{n-1}
g_{1,2}^\gamma(x,Q^2,P^2). \label{xmoment}
\ee
Using the expressions given in Eqs.(\ref{TargetMass2}-\ref{TargetMass3}) for
$g_{1,2}^\gamma(x,Q^2,P^2)$, we expand the integrands to the first order in $\kappa$. 
Then we obtain, 
\bea
g_{1}^{\gamma,n}&=& M_2^n -\kappa\frac{n^2(n+1)}{(n+2)^2}
M_2^{n+2}-\kappa\frac{4n(n+1)}{(n+2)^2}M_3^{n+2}+{\cal O}
(\kappa^2),\label{MomentG1gamma}\\
g_{2}^{\gamma,n}&=& -\frac{n-1}{n}M_2^n+\frac{n-1}{n}M_3^n 
+\kappa\frac{n(n+1)(n-1)}{(n+2)^2}M_2^{n+2}\nonumber\\
&&\hspace{2cm}-\kappa\frac{n^2(n-1)}{(n+2)^2}M_3^{n+2}
+{\cal O}(\kappa^2),\label{MomentG2gamma}
\eea
where we have used the formulas,
\be
\int_0^1 d\xi~\xi^{n-1}
H_{a,d}(\xi)=\frac{M^n_{2,3}}{n^2}~, \quad 
\int_0^1 d\xi~\xi^{n-1} G_{a,d}(\xi)=\frac{M^n_{2,3}}{n}~,\quad 
\int_0^1 d\xi~\xi^{n} F_{a,d}(\xi)=M^n_{2,3}~.
\ee
The result is consistent with the one obtained for the case of  polarized nucleon
target in Ref.\cite{PR}. For phenomenological analysis, the experimental data will be 
used for the left-hand sides of Eqs.(\ref{MomentG1gamma}-\ref{MomentG2gamma})
which should be compared with the right-hand sides, the QCD
predictions.

\section{Numerical Analysis}
\smallskip

Let us perform a numerical analysis for the target mass effects in $g_1^\gamma$ and
$g_2^\gamma$.

\subsection{$g_1^\gamma(x,Q^2,P^2)$ derived from $H_{a,d}(\xi)$, $G_{a,d}(\xi)$ and
$F_{a,d}(\xi)$}

We first compute the functions, $H_{a,d}(\xi)$, $G_{a,d}(\xi)$ 
and $F_{a,d}(\xi)$, inverting the Nachtmann moments $M_{2}^n$ and $M_{3}^n$,
\bea
H_{a,d}(\xi)&=&\frac{1}{2\pi i}\int_{c-i\infty}^{c+i\infty}
dn \,\xi^{-n} \frac{M_{2,3}^n}{n^2} , \quad \label{Hfunction}\\
G_{a,d}(\xi)&=&\frac{1}{2\pi i}\int_{c-i\infty}^{c+i\infty}
dn \,\xi^{-n}\frac{M_{2,3}^n}{n}, \quad  \label{Gfunction}\\
\xi F_{a,d}(\xi)&=&\frac{1}{2\pi i}\int_{c-i\infty}^{c+i\infty}
dn \,\xi^{-n}{M_{2,3}^n}~. \quad \label{Ffunction}
\eea
We use the QCD result for $M_2^n (= \sum_i a_{(2)i}^{\gamma,n}
E_{(2)i}^n)$, which has been calculated up to NLO and given in Eq. (3.16) of the first
article of  Ref.\cite{SU}.  As for $M_3^n (= \sum_i a_{(3)i}^{\gamma,n} E_{(3)i}^n)$, 
on the other hand, we adopt the pure QED result, Eq.(3.22) of Ref.\cite{BSU}, with 
the factor $\frac{n-1}{n}$ taken out. The QCD calculation of $M_3^n$ even in LO has not been 
accomplished yet. The evaluation of the twist-3 part $M_3^n$ in QCD is feasible when $n$ is a 
small number. But as $n$ gets larger it becomes a more and more difficult task 
due to the increase of the number of participating operators and the mixing 
among these operators \cite{Shuryaketal}.

We have plotted the twist-2 contributions, $H_{a}(\xi)$,
$G_{a}(\xi)$, and $\xi F_{a}(\xi)$ as  functions of $\xi$ in Fig. 3, and 
the twist-3 contributions, $H_{d}(\xi)$, 
$G_{d}(\xi)$, and $\xi F_{d}(\xi)$  in Fig. 4, for the case of $Q^2=30$ GeV$^2$ 
and $P^2=1$ GeV$^2$. We take $\Lambda=0.2$ GeV 
for the QCD parameter and $N_f=3$ for the number of active quark flavors 
throughout our numerical analysis.  
We see that all the functions $H_{a,d}(\xi)$, $G_{a,d}(\xi)$ and 
$F_{a,d}(\xi)$ vanish as $\xi \rightarrow 1$. 
The behavior of a function near $\xi=1$ is governed 
by its  moments at large $n$. The LO QCD result for $M_2^n$ gives 
$M_2^n \sim 1/(n {\rm ln}\!~n)$ for large $n$, 
which determines the dominant behaviors of the functions near $\xi=1$,
and thus we expect that 
\be
\xi F_a(\xi)\sim -\frac{1}{{\rm ln}(1-\xi)}~, \quad 
G_a(\xi)\sim \frac{{\rm ln}\xi}{{\rm ln}(1-\xi)}~, \quad 
H_a(\xi)\sim -\frac{({\rm ln}\xi)^2}{{\rm ln}(1-\xi)}~.
\ee
For the twist-3 part, $M_3^n$, the pure QED result tells that $M_3^n \sim -1/n^2$ at 
large $n$. So we get near $\xi=1$, 
\be
\xi F_d(\xi)\sim {\rm ln}\xi~, \quad 
G_d(\xi)\sim -({\rm ln}\xi)^2~, \quad 
H_d(\xi)\sim ({\rm ln}\xi)^3~.
\ee
The behaviors of $H_{a,d}(\xi)$, $G_{a,d}(\xi)$ and $F_{a,d}(\xi)$ 
as $\xi \rightarrow 1$ in Figs. 3 and 4 are indeed just what we have expected.
The functions $H_{a,d}(\xi)$, $G_{a,d}(\xi)$ and $F_{a,d}(\xi)$ for the case of $Q^2=10$
GeV$^2$  and $P^2=1$ GeV$^2$ show the similar  behaviors.

Putting these results into the formula (\ref{TargetMass2}) and we obtain 
$g_1^\gamma(x,Q^2,P^2)$ with TME as a function of $x$, 
which are shown (solid curve) in Fig. 5 for
$Q^2=30$ GeV$^2$ with $P^2=1$ GeV$^2$ and in Fig. 6 for $Q^2=10$ GeV$^2$ 
with $P^2=1$ GeV$^2$. The vertical axis is in units of
$3N_f\langle e^4\rangle \frac{\alpha}{\pi}\ln(Q^2/P^2)$, where
$\alpha=e^2/4\pi$, the QED coupling constant,  
and $\langle e^4\rangle =\sum_{i=1}^{N_f}e_i^4/N_f$ with 
$e_i$ being the electric charge of $i$th flavor quark.
Also plotted are  $g_1^\gamma(x,Q^2,P^2)\vert_0$ without
TME (dashed curve) (see Eq.(\ref{g1WithoutTME})) and the one with TME
included up to the first order in $P^2/Q^2$ (short-dashed curve), which is obtained 
by the inverse Mellin transform of the right-hand side of Eq.(\ref{MomentG1gamma}).
We observe that the target mass effects appear between intermediate $x$ and 
$x_{\rm max}$, and that the effects become sizable 
when the ratio $Q^2/P^2$ is reduced  (see Fig. 6). 
The distinction between the behaviors of $g_1^\gamma$ with
and without  TME is remarkable near $x_{\rm max}$. We get  
$x_{\rm max}\approx 0.97$ for $Q^2=30$ GeV$^2$ with 
$P^2=1$ GeV$^2$ and $x_{\rm max}\approx 0.91$ for $Q^2=10$ GeV$^2$ 
with $P^2=1$ GeV$^2$. The graphs of $g_1^\gamma$ with TME vanish at $x_{\rm max}$
as they should. But those graphs without TME or with TME partially included
remain finite. In the small $x$-region the target mass effects are almost negligible.
We also note that the graph with the first order corrections in $P^2/Q^2$ 
is a good approximation to the full-order result except around 
$x_{\rm max}$.

\subsection{The First Moment Sum Rule of $g_1^\gamma(x,Q^2,P^2)$ and 
Target Mass Effects}

When the target mass corrections are not taken into account, the polarized virtual 
photon structure function $g_1^\gamma(x,Q^2,P^2)$ satisfies the following sum rule \cite{NSV,SU}:
\be
\Gamma_1^\gamma\equiv\int^1_0 dx g_1^\gamma(x,Q^2,P^2)\vert_0=-\frac{3\alpha}{\pi}
\sum_{i=1}^{N_f}e_i^4+{\cal O}(\alpha_s) ~.\label{Firstg1Without}
\ee
The right-hand side corresponds to the twist-2 contribution,  
$M_2^{n=1} (=\sum_i a_{(2)i}^{\gamma,{n=1}}E_{(2)i}^{n=1})$, and 
actually the first term is the consequence of the QED axial anomaly. 
Now it will be interesting to see how this sum rule is modified when 
TME are included.

From Eq.(\ref{Nacht2}) we easily see that once the target mass corrections are
taken into account, 
the above sum rule is modified to  the first Nachtmann moment, which reads 
\bea
&&\frac{1}{9}\int_0^{x_{\rm max}}dx
\frac{\xi^2}{x^2}\left[
5+4\sqrt{1-\frac{4P^2x^2}{Q^2}}\right]g_1^\gamma(x,Q^2,P^2)\nonumber\\
&&+\frac{4}{3}\int_0^{x_{\rm max}}dx
\frac{\xi^2}{x^2}\frac{P^2x^2}{Q^2}g_2^\gamma(x,Q^2,P^2)\nonumber\\
&&\quad = -\frac{3\alpha}{\pi}\sum_{i=1}^{N_f}e_i^4+{\cal O}(\alpha_s)~.
\eea
Phenomenologically it would be appropriate to express the first moment of 
$g_1^\gamma(x,Q^2,P^2)$ itself in terms of $M_2^n$ and $M_3^n$, which are 
calculable by perturbative QCD. Setting $n=1$ in Eq.(\ref{MomentG1gamma}), we obtain 
to the first order in $P^2/Q^2$, 
\bea
&&\int_0^{x_{\rm max}}dx g_1^\gamma(x,Q^2,P^2)= M_2^{n=1} - \left\{\frac{2}{9}
M_2^{n=3}+\frac{8}{9}M_3^{n=3}\right\}\frac{P^2}{Q^2}+{\cal O}
\Bigl((P^2/Q^2)^2 \Bigr) ~.\label{Firstg1With}\nn\\
\eea
where $M_2^{n=1}= -(3\alpha/\pi)\sum_{i=1}^{N_f}e_i^4$. Thus the target mass corrections
$\Delta \Gamma_1^\gamma$ to the first moment of $g_1^\gamma$, i.e., the difference 
between the left-hand sides of  Eq.(\ref{Firstg1With}) and (\ref{Firstg1Without}) is given,
to the first order in $P^2/Q^2$, by
\be
\Delta \Gamma_1^\gamma=- \left\{\frac{2}{9}
M_2^{n=3}+\frac{8}{9}M_3^{n=3}\right\}\frac{P^2}{Q^2}~.
\ee
Up to this order in $P^2/Q^2$ we only need to know the reduced matrix elements and 
coefficient functions for $n=3$.   Using the NLO result for
$M_2^{n=3}$ in QCD, given   in Eq.(3.16) of Ref.\cite{SU}, we obtain 
\be
M_2^{n=3}/\frac{\alpha}{\pi}=0.163~ (0.0601)
\quad \mbox{for} \quad Q^2=30~{\rm GeV}^2(10~{\rm GeV}^2),\quad P^2=1{\rm GeV}^2~.\nonumber\\
\label{M2n3}
\ee
As for the twist-3 contribution $M_3^{n=3}$, the LO result in QCD is available. 
Taking the results in Eqs.(4.30-4.36) of Ref.\cite{BSU}, we get
\be
M_3^{n=3}/\frac{\alpha}{\pi}=-0.130~ (-0.0942)
\quad \mbox{for} \quad Q^2=30~{\rm GeV}^2(10~{\rm GeV}^2),\quad P^2=1{\rm GeV}^2~.\nonumber\\
\label{M3n3}
\ee
With these numerical values we find
\be
\Delta \Gamma_1^\gamma/M_2^{n=1}=-0.00395(-0.0106)\quad \mbox{for} \quad Q^2=30~
{\rm GeV}^2(10~{\rm GeV}^2),\quad P^2=1{\rm GeV}^2~.\nonumber\\
\ee

The target mass corrections to the first moment sum rule
of $g_1^\gamma$ amount to $0.40\%$ ($1.1\%$) for $Q^2=30$GeV$^2$ ($10$GeV$^2$) 
with $P^2=1$GeV$^2$, which are negligibly small. Even for the latter case, 
$Q^2=10$GeV$^2$ and $P^2=1$GeV$^2$, the corrections are, at most, of order of one percent.
We see from Eqs.(\ref{M2n3}-\ref{M3n3}) that the twist-2 and twist-3 contributions $M_2^{n=3}$
and $M_3^{n=3}$ for $n=3$ are almost the same in magnitude but have the opposite signs. This
leads to the smallness of  target mass corrections to the first moment sum rule of $g_1^\gamma$. 

\subsection{$g_2^\gamma(x,Q^2,P^2)$ and the Target Mass Effects}

We obtain the graph of $g_2^\gamma(x,Q^2,P^2)$ with TME 
by inserting the functions $H_{a,d}(\xi)$, $G_{a,d}(\xi)$ and $F_{a,d}(\xi)$ derived 
from Eqs.(\ref{Hfunction}-\ref{Ffunction}) into Eq.(\ref{TargetMass3}). Again we have used 
the pure QED result for $M_3^n$, Eq.(3.22) of Ref.\cite{BSU},  since the
QCD result  for $M_3^n$ with $n> 3$ is not available. In Fig. 7, we
have plotted $g_2^\gamma(x,Q^2,P^2)$ with TME (solid curve) in units 
of $3N_f\langle e^4\rangle \frac{\alpha}{\pi}\ln(Q^2/P^2)$, for
$Q^2=30$ GeV$^2$ and $P^2=1$ GeV$^2$. Also shown in Fig. 7 is the
box-diagram  contribution  to $g_2^\gamma$ (dashed curve)  for an example
without TME, the expression of which is given by~\cite{BSU}
\be
g_2^{\gamma({\rm Box})}(x,Q^2,P^2)=
\frac{3\alpha}{\pi}N_f\langle e^4 \rangle
\left[-(2x-1)\ln{\frac{Q^2}{P^2}}+2(2x-1)\ln{x}+6x-4\right]~.
     \label{g2gamma}
\ee
The graph of $g_2^\gamma$ with TME vanishes at $x_{\rm max}$, 
but not the one without TME. 

In a certain limit the analysis of $M_3^n$ in QCD becomes tractable. 
The contribution to $M_3^n$ is made up of two components;
the flavor singlet and nonsinglet. In an approximation of
neglecting terms of order ${\cal O}(1/N_c^2)$,  we are able to calculate
$M_3^{n({\rm NS})}$, the flavor nonsinglet contribution to $M_3^n$, for 
arbitrary $n$ in QCD since in this limit the problem of 
operator mixing  can be evaded~\cite{ABHetal}. In fact,  we have computed
$M_3^{n({\rm NS})}$ in LO QCD for large $N_c$  limit, which is given in
Eq.(4.40) of  Ref.\cite{BSU}. Using these $M_3^{n({\rm NS})}$ we  perform
the inverse Mellin transform  of Eqs.(\ref{Hfunction}-\ref{Ffunction}) to
obtain the flavor nonsinglet contributions, 
$H^{\rm NS}_d(\xi)$, $G^{\rm NS}_d(\xi)$ and $F^{\rm NS}_d(\xi)$.   
Then putting these functions into the formula (\ref{TargetMass3}) and 
setting the twist-2 contributions to zero, i.e., $H_a(\xi)=G_a(\xi)=F_a(\xi)=0$, 
we obtain ${\overline g_2}^{\gamma({\rm NS})}(x,Q^2,P^2)$, the 
twist-3 contribution to the flavor nonsinglet part of $g_2^\gamma$, including TME. 

In Fig. 8, we have plotted ${\overline g_2}^{\gamma({\rm NS})}(x,Q^2,P^2)$ with TME 
(solid curve)  in units of $3N_f(\langle e^4\rangle-\langle
e^2\rangle^2)\frac{\alpha}{\pi}\ln(Q^2/P^2)$, for $Q^2=30$  GeV$^2$ and
$P^2=1$ GeV$^2$, where $(\langle e^4\rangle-\langle
e^2\rangle^2)$ is a charge factor for the flavor nonsinglet component 
with $\langle e^2\rangle =\sum_{i=1}^{N_f}e_i^2/N_f$. 
Also plotted are the graphs of ${\overline g_2}^{\gamma({\rm NS})}$ without 
TME (short-dashed curve) and the box-diagram  contribution  to 
${\overline g_2}^{\gamma({\rm NS})}$ 
(dashed curve) whose expression is given  by~\cite{BSU}
\bea
{\overline g}_2^{\gamma({\rm NS, Box})}
&=&\frac{3\alpha}{\pi}N_f(\langle e^4\rangle-\langle e^2\rangle^2) \nn\\
&& \times\Bigl[(2x-2-\ln{x})\ln{\frac{Q^2}{P^2}}
-2(2x-1)\ln{x}+2(x-1)+{\ln}^2{x}\Bigr].\label{g2gammaBoxx}
\eea
We observe that  target mass corrections in the twist-3 part are negligibly
small. This is inferred from the fact that target mass effects appear
at large 
$\xi$ (large $x$) and 
the twist-3 contributions 
$\xi F_d(\xi)$, $G_d(\xi)$, and $H_d(\xi)$ vanish as ${\rm ln}\xi$, 
$-({\rm ln}\xi)^2$, and $({\rm ln}\xi)^3$, respectively, for $\xi \rightarrow 1$.
Another case for $Q^2=10$ GeV$^2$  with $P^2=1$ GeV$^2$ is shown in Fig. 9, 
where we see that target mass effects 
become slightly larger than the case for $Q^2=30$ GeV$^2$, in particular, in
the region near $x_{\rm max}$.

\subsection{Burkhardt-Cottingham Sum Rule}

Just as the spin-dependent nucleon structure function $g_2^{\rm nucl}$, 
the polarized virtual photon structure function $g_2^\gamma(x,Q^2,P^2)$ satisfies 
the Burkhardt-Cottingham (BC) sum rule \cite{BC}:
\be
\int^1_0 dx g_2^\gamma(x,Q^2,P^2)\vert_0=0~. \label{BCWithout}
\ee
We put the subscript 0 to emphasize that this is the statement when the target mass corrections 
are not included. When the TME are included, Eq.(\ref{MomentG2gamma}) shows that 
the BC sum rule still holds up to the first order in $P^2/Q^2$. Actually we take the $x$-moments of
$g_2^\gamma$ whose expression is given in Eq.(\ref{TargetMass3}). Using the relations in 
Eq.(\ref{GFad}) and by partial integration with the boundary conditions 
$H_{a,d}(\xi=1)=G_{a,d}(\xi=1)=0$, we obtain 
\bea
\int_0^{x_{\rm max}}dx{x}^{n-1}g_2^\gamma(x,Q^2,P^2)&=&n(n-1)\int^1_0d\xi
\frac{\xi^{n-1}}{(1+\kappa\xi^2)^{n+1}} \nn \\
&&\qquad \qquad \times\Bigl[-H_a(\xi)+(1+ \kappa\xi^2)H_d(\xi)  \Bigr]~. \label{xmomentg2gamma}
\eea
Taking $n=1$, we arrive at
\be
\int_0^{x_{\rm max}}dx \ g_2^\gamma(x,Q^2,P^2)=0~,
\ee
which shows the BC sum rule is free from the target mass effects. Note that 
the upper limit of integration has changed from 1 to $x_{\rm max}$.
A similar expression to the one in Eq.(\ref{xmomentg2gamma}) has been 
obtained by Piccione and Ridolfi \cite{PR} for the moments of the 
nucleon structure function $g_2^{\rm nucl}$ when the target mass corrections are included.

\section{Conclusion}
\smallskip
In this paper, we have investigated the target mass effects 
in the polarized virtual photon structure functions 
$g_1^\gamma (x,Q^2,P^2)$, $g_2^\gamma (x,Q^2,P^2)$ which can be 
measured in the future experiments of the polarized version of 
the $ep$ or $e^+e^-$ colliders. 

We based our argument on the framework of the OPE and the O(4) expansion,
taking into account the trace terms of the operators of the definite
spin. This amounts to the use of the expansion of the amplitudes 
in terms of the Gegenbauer polynomials and their orthogonality relations 
to extract the contributions with the definite spin, given as
the Nachtmann moments. The evaluation of the \lq\lq kinematical\rq\rq \ 
target mass effects is important to extract the \lq\lq dynamical\rq\rq\ 
higher-twist 
effects which would also exist in the power corrections in $P^2/Q^2$.
  
We have derived the expressions for $g_1^\gamma (x,Q^2,P^2)$ 
and $g_2^\gamma (x,Q^2,P^2)$ in closed form by inverting the
Nachtmann moments for the twist-2 and twist-3 operators.
The characteristic feature for the photon target compared to
the nucleon case is the presence of the maximal value $x_{\rm max}(<1)$
of the Bjorken variable $x$, while $\xi(x_{\rm max})=1$. Hence we do not 
encounter the similar problem due to the kinematical relation 
$\xi(x_{\rm max})<1$ with $x_{\rm max}=1$ in the case of the nucleon. 

Our numerical analysis shows that the target mass effects 
appear at large $x$ and become sizable near $x_{\rm max}(=1/(1+\frac{P^2}{Q^2}))$, as the ratio $P^2/Q^2$ increases. The structure functions
evaluated by inverting the Nachtmann moments in fact vanish at $x=x_{\rm max}$.
We have also examined the target mass effects for the first-moment sum rules of $g_1^\gamma$ and $g_2^\gamma$. For the kinematic region we consider, the corrections to the first moment of $g_1^\gamma$ turn out to be negligibly small. The 
first moment of $g_2^\gamma$ leads to the Burkhardt-Cottingham sum rule, where
the upper limit of integration becomes $x_{\rm max}$.
More thorough QCD analysis including the flavor-singlet part of $g_2^\gamma$
is now under investigation.


\vspace{0.5cm}

\leftline{\large\bf Acknowledgements}

\vspace{0.5cm}

This work is partially supported by the
Grant-in-Aid for Scientific Research from the Ministry
of Education, Culture, Sports, Science and Technology, NO.(C)(2)-15540266.


\newpage

\appendix


\noindent

{\LARGE\bf Appendix}


\bigskip

In this appendix we present main formulas of Gegenbauer polynomials which are  
used in this paper and give an outline of the derivation of Eqs.(\ref{cont-tw2}-\ref{cont-tw3}).

\section{Gegenbauer polynomials}
The Gegenbauer polynomials $C_n^{(\nu)}(\eta)$ are defined through the generating 
function given by
\bea
(1-2\eta t+t^2)^{-\nu}=\sum_{n=0}^\infty C_n^{(\nu)}(\eta)t^n.
\eea
In terms of hypergeometric functions $F(\alpha,\beta,\gamma; z)$, $C_n^{(\nu)}(\eta)$ is
expressed as
\bea
C_n^{(\nu)}(\eta)&=&\frac{2^n \Gamma(n+\nu)}{n!\Gamma(\nu)}F\Bigl(-\frac{n}{2}, 
\frac{1-n}{2},1-n-\nu; \frac{1}{\eta^2}  \Bigr)\nn\\
&=&\frac{1}{\Gamma(\nu)}\sum_{j=0}^{n/2}
\frac{(-1)^j~\Gamma(\nu+n-j)}{j!~(n-2j)!}(2\eta)^{n-2j}~,
\eea
from which we obtain, for example, 
\bea
C_n^{(1)}(\eta)&=&\sum_{j=0}^{n/2} \frac{(-1)^j}{j!}\frac{(n-j)!}{(n-2j)!}
(2\eta)^{n-2j}~,  \label{Cn1}\\
C_{n-1}^{(2)}(\eta)&=&\sum_{j=0}^{(n-1)/2}
\frac{(-1)^j}{j!}\frac{(n-j)!}{(n-2j-1)!}\, (2\eta)^{n-2j-1} \label{exp}~.
\eea

\subsection{Recursion formulas}
\bea
&&nC_n^{(\nu)}(\eta)=2\nu[\eta C_{n-1}^{(\nu+1)}(\eta)-C_{n-2}^{(\nu+1)}(\eta)]~,
\label{Recursion1}\\
&&(n+2\nu)C_n^{(\nu)}(\eta)=2\nu[C_n^{(\nu+1)}(\eta)-
\eta C_{n-1}^{(\nu+1)}(\eta)]~,\label{Recursion2}\\
&&(n+2)C_{n+2}^{(\nu)}(\eta)=2(n+\nu+1)\eta C_{n+1}^{(\nu)}(\eta)
-(n+2\nu)C_n^{(\nu)}(\eta)~.\label{Recursion3}
\eea

\subsection{Orthogonality relations}
\bea
\int_{-1}^{1}(1-\eta^2)^{\nu-\frac{1}{2}}C_m^{(\nu)}(\eta)C_n^{(\nu)}(\eta)d\eta
=\frac{2\pi}{2^{2\nu}}\frac{\Gamma(n+2\nu)}{(n+\nu)n![\Gamma(\nu)]^2}
\delta_{mn}.\label{orthogonal}
\eea
In addition we need 
the following formula for the integral to project out the 
contributions  of definite spin from the dispersion relations,
\bea
&&\int_{-1}^{1}d\eta \ \eta^m (1-\eta^2)^{\nu-1/2}C_n^{(\nu)}(\eta)\frac{1}
{\zeta-\eta}=\frac{\pi}{2^{\nu-1}} \zeta^m(\zeta^2-1)^{\frac{\nu-1}{2}}\left[
\zeta-(\zeta^2-1)^{1/2}\right]^{n+\nu}\nonumber\\
&&\hspace{2cm}\times\frac{\Gamma(n+2\nu)}{\Gamma(\nu)\Gamma(n+\nu+1)} 
F\left(\nu,1-\nu,n+\nu+1;\frac{-\zeta+(\zeta^2-1)^{1/2}}{2(\zeta^2-1)^{1/2}}
\right). \label{IntegralGegenbauer}
\eea
In fact the factor $\left[\zeta-(\zeta^2-1)^{1/2}\right]^{n+\nu}$ gives 
$\left(-\frac{P}{Q}\right)^{n+\nu}~\xi^{n+\nu}$, where $\xi$ is the 
so-called $\xi$-scaling variable given in Eq.(\ref{xi}).

\section{Derivation of Eqs.(\ref{cont-tw2}-\ref{cont-tw3})}

\subsection{Contraction formulas}

We give a table of contractions which are used for the derivation. 
First we introduce the most general rank-$n$ symmetric and traceless tensor, 
$\Pi^{\mu_1\cdots \mu_{n}}$,  that can be formed  with the momentum $p$ alone~\cite{GP}, 
\be
\Pi^{\mu_1\cdots \mu_{n}}=\sum_{j=0}^{n/2}\frac{(-1)^j}{2^j}\frac{(n-j)!}{n!}
\underbrace{g\cdots g}_j \overbrace{p\cdots p}^{n-2j}~(p^2)^j ~, 
\ee
where $\displaystyle{\underbrace{g\cdots g}_j}$ stands for a product of
$j$ metric tensors $g^{\mu_l\mu_k}$ with $2j$ indices chosen among
$\mu_1,\cdots,\mu_{n}$ in all possible ways. Then we easily find that 
the contraction of $\Pi^{\mu_1\cdots \mu_{n}}$ 
with $q_{\mu_1}\cdots q_{\mu_{n}}$ is expressed in terms of 
Gegenbauer polynomial $C_n^{(1)}(\eta)$ given in Eq.(\ref{Cn1}) 
\cite{NACHT,WAND},
\be
I_{(0)}(n)\equiv q_{\mu_1}\cdots q_{\mu_{n}}\Pi^{\mu_1\cdots \mu_{n}}
=a^n C_n^{(1)}(\eta)~,  \label{Izero}
\ee
where
\be
a=-\frac{1}{2}PQ, \quad  \eta=-\frac{p\cdot q}{PQ}\label{variable}~. 
\ee

Next we differentiate the both sides of Eq.(\ref{Izero}) with respect to $p^\beta$. 
Using the following formulas, 
\be
\frac{\partial a}{\partial p^\beta}=a~\Bigl(\frac{-p_\beta}{P^2}  \Bigr)~, 
\qquad \frac{\partial \eta}{\partial p^\beta}=\frac{q_\beta}{2a}+
\eta\frac{p_\beta}{P^2}, \qquad 
\frac{dC_n^{(\nu)}(\eta)}{d\eta}=2\nu C_{n-1}^{(\nu+1)}(\eta)~, \label{DerivETA}\ee
we find 
\be
n~ q_{\mu_1}\cdots q_{\mu_{n}}\{\delta_\beta^{\mu_1} p^{\mu_2}\cdots p^{\mu_{n-1}}\}_n=\left\{-
na^{n}\frac{p_\beta}{P^2}C_n^{(1)}(\eta)+ a^n
\Bigl[\frac{q_\beta}{2a}+\eta\frac{p_\beta}{P^2}\Bigr]2 C_{n-1}^{(2)}(\eta)\right\}~,\\
\ee
where
$\{\delta_\beta^{\mu_1} p^{\mu_2}\cdots p^{\mu_{n-1}}\}_n$ is a tensor which 
is formed with one $\delta_\beta^{\mu_i}$ and $n\!-\!1$ momentum four-vectors $p$
and totally symmetric among indices $\mu_1\cdots \mu_{n}$. Moreover, it is traceless 
in the sence that $g_{\mu_i \mu_j}\{\delta_\beta^{\mu_1} p^{\mu_2}\cdots p^{\mu_{n-1}}\}_n=0$ 
for all pairs $i$, $j$. 
Now it is reminded that 
the polarized photon matrix elements are multiplied by the factor
${\epsilon_{\rho\tau\alpha}}^\beta$. Thus the terms with $p_\beta$ in contractions do not
contribute  in the end. Also the terms with $q^\sigma$ 
which appear later on give null results when multiplied by  
${\epsilon_{\mu\nu\lambda\sigma}}^\lambda$ (see Eq.(\ref{Amplitude})). So we obtain
\be
I_{(1)\beta}(n)=q_{\mu_1}\cdots q_{\mu_{n}}\{\delta_\beta^{\mu_1} 
p^{\mu_2}\cdots p^{\mu_{n-1}}\}_n =\frac{1}{n}q_\beta a^{n-1}C_{n-1}^{(2)}(\eta)
+({\rm terms\ with}\ p_\beta)~.\label{Ione}
\ee

Further we differentiate the both sides of Eq.(\ref{Ione}) with respect to $q_\sigma$. 
With 
\be
\frac{\partial a}{\partial q_\sigma}=a~\Bigl(\frac{-q^\sigma}{Q^2}  \Bigr)~, 
\qquad \frac{\partial \eta}{\partial q_\sigma}=\frac{p^\sigma}{2a}+
\eta\frac{q^\sigma}{Q^2}~,
\ee
we obtain
\bea
&&I_{(2)\beta}^\sigma(n-1)=q_{\mu_1}\cdots q_{\mu_{n-1}}\{\delta_\beta^{\sigma} p^{\mu_1}\cdots p^{\mu_{n-1}}\}_n =\frac{1}{n}\frac{\partial I_{(1)\beta}(n)}{\partial q_\sigma}
\nn\\
&&\  =\frac{1}{n^2}\Bigl\{{\delta_\beta}^\sigma a^{n-1}C_{n-1}^{(2)}(\eta)
+q_\beta p^\sigma a^{n-2}2C_{n-2}^{(3)}(\eta) \Bigr\}
+({\rm terms\ with}\ p_\beta\  {\rm or}\ q^\sigma). \hspace{1cm}\label{Itwo}
\eea
Finally the both sides of Eq.(\ref{Izero}) are differentiated with respect to $q_\sigma$, 
and we  get
\bea
I_{(3)}^\sigma(n-1)&=&q_{\mu_1}\cdots q_{\mu_{n-1}}\Pi^{\sigma\mu_1 \cdots \mu_{n-1}}
=\frac{1}{n}\frac{\partial I_{(0)}(n)}{\partial q_\sigma} \nn\\
&=&\frac{1}{n}\Bigl\{p^\sigma a^{n-1}C_{n-1}^{(2)}(\eta)\}
+({\rm terms\ with}\  q^\sigma).\label{Ithree}
\eea
The terms with $q^\sigma$ which appear in Eqs.(\ref{Itwo}-\ref{Ithree}) have been omitted.
With Eqs.(\ref{Izero}, \ref{Ione}, \ref{Itwo}, \ref{Ithree}) at hand,  we are now ready to derive 
Eqs.(\ref{cont-tw2}-\ref{cont-tw3}).

\subsection{Derivation}

The tensor ${{\widetilde M}_{(2)\beta}}^{\sigma\mu_1\cdots\mu_{n-1}}$, which  
corresponds to the traceless twist-2 operator $R^n_{(2)i}$, is formed one 
$\delta_\beta^\sigma$ (or $\delta_\beta^{\mu_i}$)
and $n\!-\!1$ momentum four-vectors $p$
and totally symmetric among indices $\sigma, \mu_1\cdots \mu_{n-1}$. Thus it is given by
\be
{{\widetilde M}_{(2)\beta}}^{\sigma\mu_1\cdots\mu_{n-1}}=
\{{\delta_\beta}^\sigma p^{\mu_1}\cdots p^{\mu_{n-1}}\}_n~.
\ee
For its contraction with $q_{\mu_1}\cdots q_{\mu_{n-1}}$, we find from Eq.(\ref{Itwo})
\bea
&&q_{\mu_1}\cdots q_{\mu_{n-1}}{{\widetilde M}_{(2)\beta}}^{\sigma\mu_1\cdots\mu_{n-1}}
=I_{(2)\beta}^\sigma(n-1) \nn\\
&& =\frac{1}{n^2}\Bigl\{{\delta_\beta}^\sigma a^{n-1}C_{n-1}^{(2)}(\eta)
+q_\beta p^\sigma a^{n-2}2C_{n-2}^{(3)}(\eta) \Bigr\}
+({\rm terms\ with}\ p_\beta\  {\rm or}\ q^\sigma).\hspace{1cm}
\eea

The tensor ${{\widetilde M}_{(3)\beta}}^{\sigma\mu_1\cdots\mu_{n-1}}$, which  
corresponds to the traceless twist-3 operator $R^n_{(3)i}$, is also formed one 
$\delta_\beta^\sigma$ (or $\delta_\beta^{\mu_i}$)
and $n\!-\!1$ momentum four-vectors $p$. Its indices 
are totally symmetric among $\mu_1\cdots\mu_{n-1}$ but
antisymmetric under $\sigma \leftrightarrow \mu_i$.
The one which satisfies these requirements is~\cite{PR}
\bea
{{\widetilde M}_{(3)\beta}}^{\sigma\mu_1\cdots\mu_{n-1}}
&=&
\frac{n-1}{n}\biggl\{\frac{n+1}{n}\Bigl[{\delta_\beta}^{\sigma}\Pi^{\mu_1 
\cdots
\mu_{n-1}} - \frac{1}{(n-1)}\sum_{l=1}^{n-1}{\delta_\beta}^{\mu_l}
\Pi^{\sigma\mu_1 \cdots 
(\mu_l)\cdots\mu_{n-1}} \Bigr]\nonumber\\ 
&& \qquad \quad +\frac{n-1}{n}\Bigl[p^{\sigma}{{\widetilde
M}_{(2)\beta}}^{\mu_1\cdots\mu_{n-1}} -\frac{1}{(n-1)}\sum_{l=1}^{n-1}
p^{\mu_l}{{\widetilde M}_{(2)\beta}}^{\sigma\mu_1\cdots 
(\mu_l)\cdots\mu_{n-1}} \Bigr]\biggr\}~.\nn\\
\eea
Using Eqs.(\ref{Izero}, \ref{Ione}, \ref{Itwo}, \ref{Ithree}), we obtain
\bea
&&q_{\mu_1}\cdots q_{\mu_{n-1}}
{{\widetilde M}_{(3)\beta}}^{\sigma\mu_1\cdots\mu_{n-1}} 
=\frac{n\!-\!1}{n}\biggl\{\frac{n\!+\!1}{n}\Bigl[{\delta_\beta}^{\sigma}I_{(0)}(n\!-\!1)
- q_\beta {I_{(3)}}^\sigma (n\!-\!2) \Bigr]\nonumber\\ 
&& \hspace{5.5cm} +\frac{n-1}{n}\Bigl[p^{\sigma}{I_{(1)}}_\beta
(n\!-\!1)-p\cdot q{I_{(2)}}_\beta^\sigma (n-2)\Bigr]\biggr\} \nn\\
&&=\frac{1}{n^2}{\delta_\beta}^{\sigma}a^{n-1}\Bigl\{ 
 (n\!-\!1)(n\!+\!1)C_{n-1}^{(1)}(\eta)-2\eta C_{n-2}^{(2)}(\eta)\Bigr\}\nn\\
&& \qquad  +\frac{1}{n^2}q_\beta p^\sigma a^{n-2}\Bigl\{ 
- 2C_{n-2}^{(2)}(\eta) -4\eta C_{n-3}^{(3)}(\eta)\Bigr\} 
+({\rm terms\ with}\ p_\beta\  {\rm or}\ q^\sigma)\nn\\
&&=\frac{1}{n^2}{\delta_\beta}^{\sigma}a^{n-1}
\Bigl\{ (n\!-\!1)C_{n-1}^{(2)}(\eta)-(n\!+\!1) C_{n-3}^{(2)}(\eta)\Bigr\}\nn\\
&&\qquad-\frac{1}{n^2}q_\beta p^\sigma a^{n-2}2\Bigl\{ 
C_{n-2}^{(3)}(\eta)+C_{n-4}^{(3)}(\eta)\Bigr\}+({\rm terms\ with}\ p_\beta\  {\rm or}\ q^\sigma)~,
\eea
where at the final stage the recursion relations in Eqs.(\ref{Recursion1}-\ref{Recursion2})
were used.


\newpage



\newpage

\vspace{2cm}

\noindent

{\large Figure Captions}

\baselineskip 16pt

\begin{enumerate}
\item[Fig. 1] \quad
Deep inelastic scattering on a polarized virtual photon
in polarized $e^{+}e^{-}$ collision,
$e^{+}e^{-} \rightarrow  e^{+}e^{-} + $ hadrons (quarks and gluons).
The arrows indicate the polarizations of the $e^{+}$ and $e^{-}$.
The mass squared of the \lq\lq probe\rq\rq
\ (\lq\lq target\rq\rq)
photon is $-Q^2(-P^2)$ ($\Lambda^2 \ll P^2 \ll Q^2$) with $\Lambda$
being the QCD scale parameter.

\item[Fig. 2] \quad
Forward scattering of a virtual photon with momentum $q$ and
another virtual photon with momentum $p$. The Lorentz indices are
denoted by $\mu,\nu,\rho,\tau$.

\item[Fig. 3] \quad
The functions $H_a(\xi)$ (solid curve) , $G_a(\xi)$ (dash-dotted curve), 
$\xi F_a(\xi)$ (dashed curve) obtained by the
inverse Mellin transform of the weighted  moments of $M_2^n$ for
twist-2 operators, Eqs.(\ref{Hfunction}), (\ref{Gfunction}), and 
(\ref{Ffunction}).

\item[Fig. 4]\quad
The functions $H_d(\xi)$ (solid curve) , $G_d(\xi)$ (dash-dotted curve), 
$\xi F_d(\xi)$ (dashed curve) obtained by the
inverse Mellin transform of the weighted moments of $M_3^n$ for
twist-3 operators, Eqs.(\ref{Hfunction}), (\ref{Gfunction}), and 
(\ref{Ffunction}).

\item[Fig. 5]\quad
The graphs of $g_1^\gamma(x,Q^2,P^2)$ with full TME
(Eq.(\ref{TargetMass2}), solid curve), with the first order TME
(short-dashed curve) and without TME (dashed curve) in units of
$3N_f\langle e^4\rangle \frac{\alpha}{\pi}\ln(Q^2/P^2)$, 
for $Q^2=30$ GeV$^2$ and $P^2=1$ GeV$^2$ with $\Lambda=0.2$ GeV, $N_f=3$.

\item[Fig. 6]\quad
The graphs of $g_1^\gamma(x,Q^2,P^2)$ 
for $Q^2=10$ GeV$^2$ and $P^2=1$ GeV$^2$ with $\Lambda=0.2$ GeV, $N_f=3$.

\item[Fig. 7]\quad
The graph of $g_2^\gamma(x,Q^2,P^2)$ with TME (Eq.(\ref{TargetMass3}),
solid curve) and the box-diagram contribution to $g_2^\gamma$ 
(dashed curve)  in units of $3N_f\langle e^4\rangle
\frac{\alpha}{\pi}\ln(Q^2/P^2)$,  for $Q^2=30$ GeV$^2$ and $P^2=1$
GeV$^2$ with $\Lambda=0.2$ GeV, $N_f=3$. 

\item[Fig. 8]\quad
The graphs of ${\overline g_2}^{\gamma({\rm NS})}(x,Q^2,P^2)$, 
the twist-3 contribution to the flavor nonsinglet part of 
$g_2^\gamma$, in units of $3N_f(\langle e^4\rangle-\langle
e^2\rangle^2)\frac{\alpha}{\pi}\ln(Q^2/P^2)$, for
$Q^2=30$  GeV$^2$ and $P^2=1$ GeV$^2$ with $\Lambda=0.2$ GeV, $N_f=3$.
The twist-3 effects are  evaluated in LO QCD for large $N_c$ limit.
The solid and short-dashed curves show the results with TME and 
without TME, respectively. We have also shown the box-diagram 
contribution to ${\overline g_2}^{\gamma({\rm NS})}$ 
(dashed curve) for comparison.  

\item[Fig. 9]\quad
The graphs of ${\overline g_2}^{\gamma({\rm NS})}(x,Q^2,P^2)$ for
$Q^2=10$  GeV$^2$ and $P^2=1$ GeV$^2$ with $\Lambda=0.2$ GeV, $N_f=3$.

\end{enumerate}

\pagestyle{empty}
\input epsf.sty
\begin{figure}
\vspace*{2cm}
\centerline{
\epsfxsize=11cm
\epsfbox{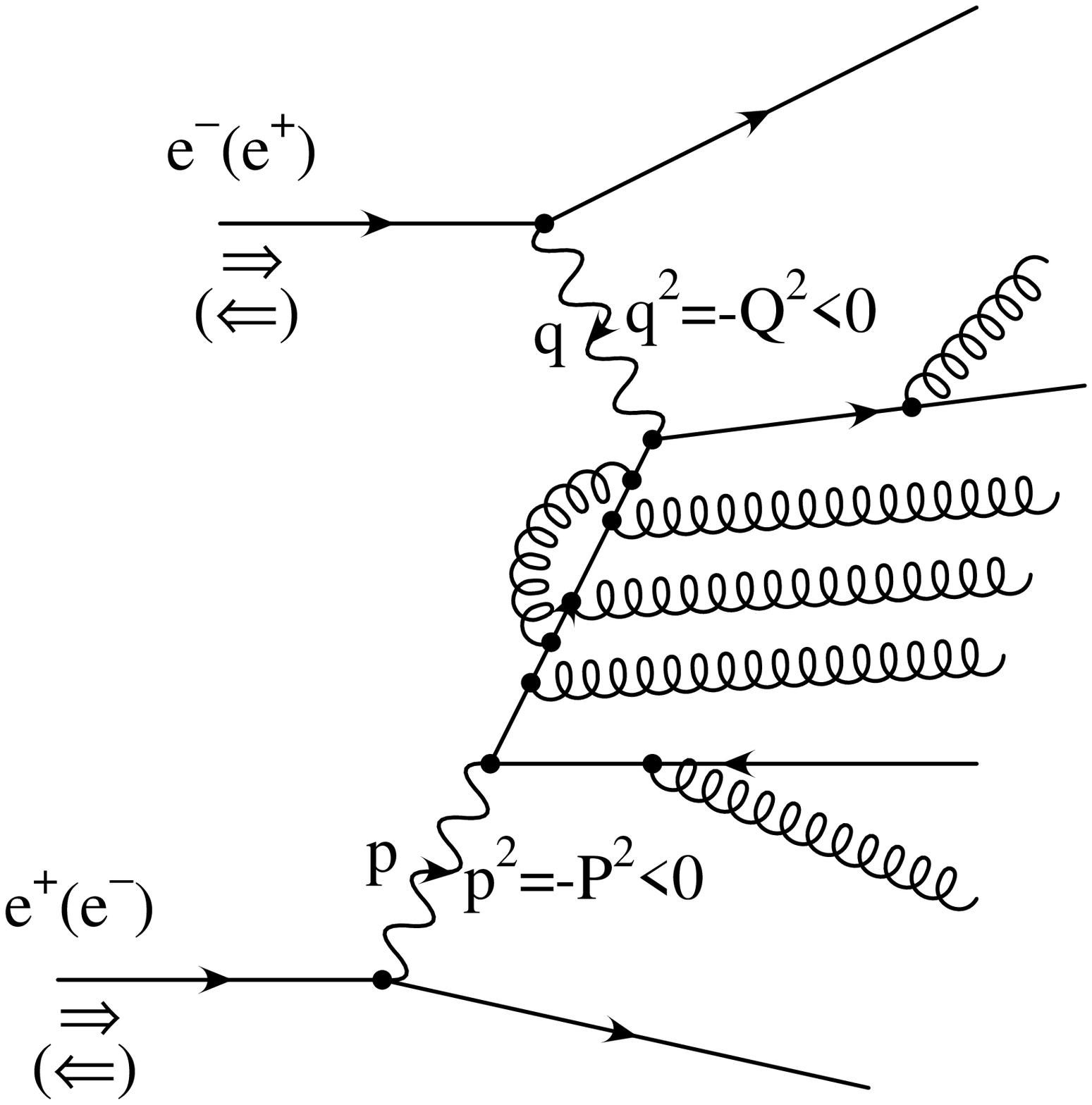}}
\vspace{-1.5cm}
\vspace{2cm}
\centerline{\large\bf Fig. 1}
\end{figure}

\newpage
\pagestyle{empty}
\input epsf.sty
\begin{figure}
\vspace*{3.5cm}
\centerline{
\epsfxsize=8cm
\epsfbox{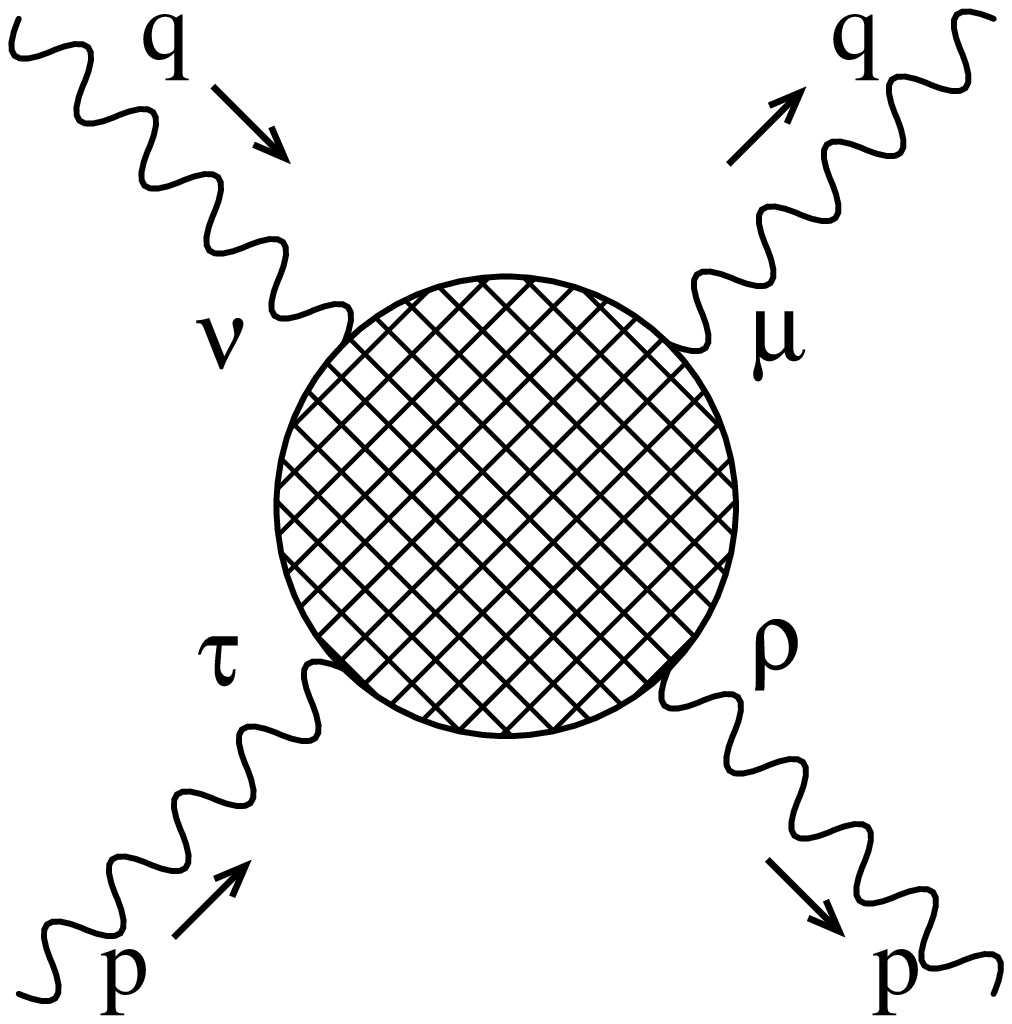}}
\vspace*{1cm}
\centerline{\large\bf Fig. 2}
\end{figure}

\newpage
\pagestyle{empty}
\input epsf.sty
\begin{figure}
\vspace*{1cm}
\centerline{
\epsfxsize=16cm
\vspace*{2cm}
\epsfbox{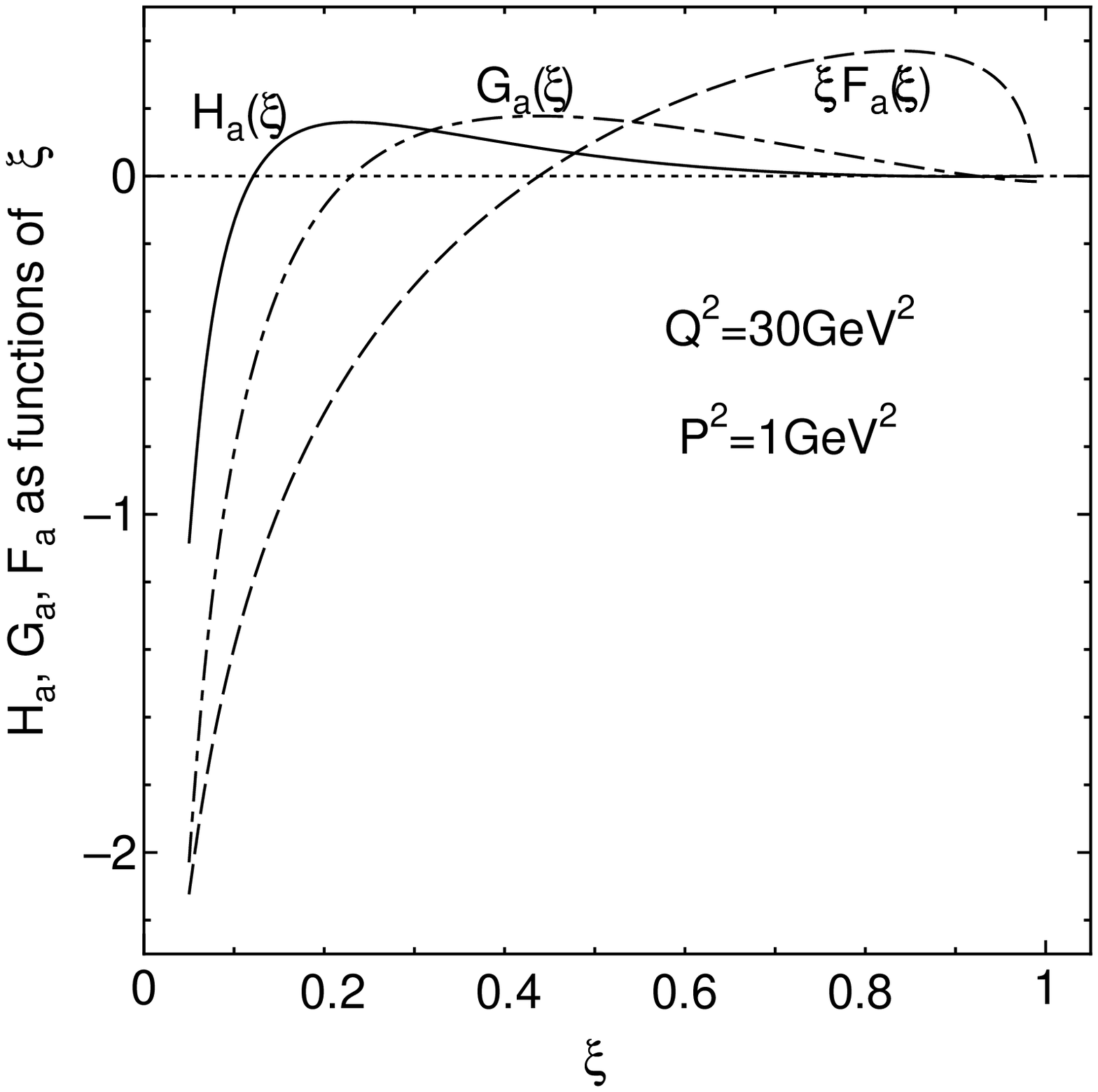}}
\centerline{\large\bf Fig. 3}
\end{figure}

\newpage
\pagestyle{empty}
\input epsf.sty
\vspace*{-2cm}
\begin{figure}
\begin{center}
\epsfxsize=16cm
\epsfbox{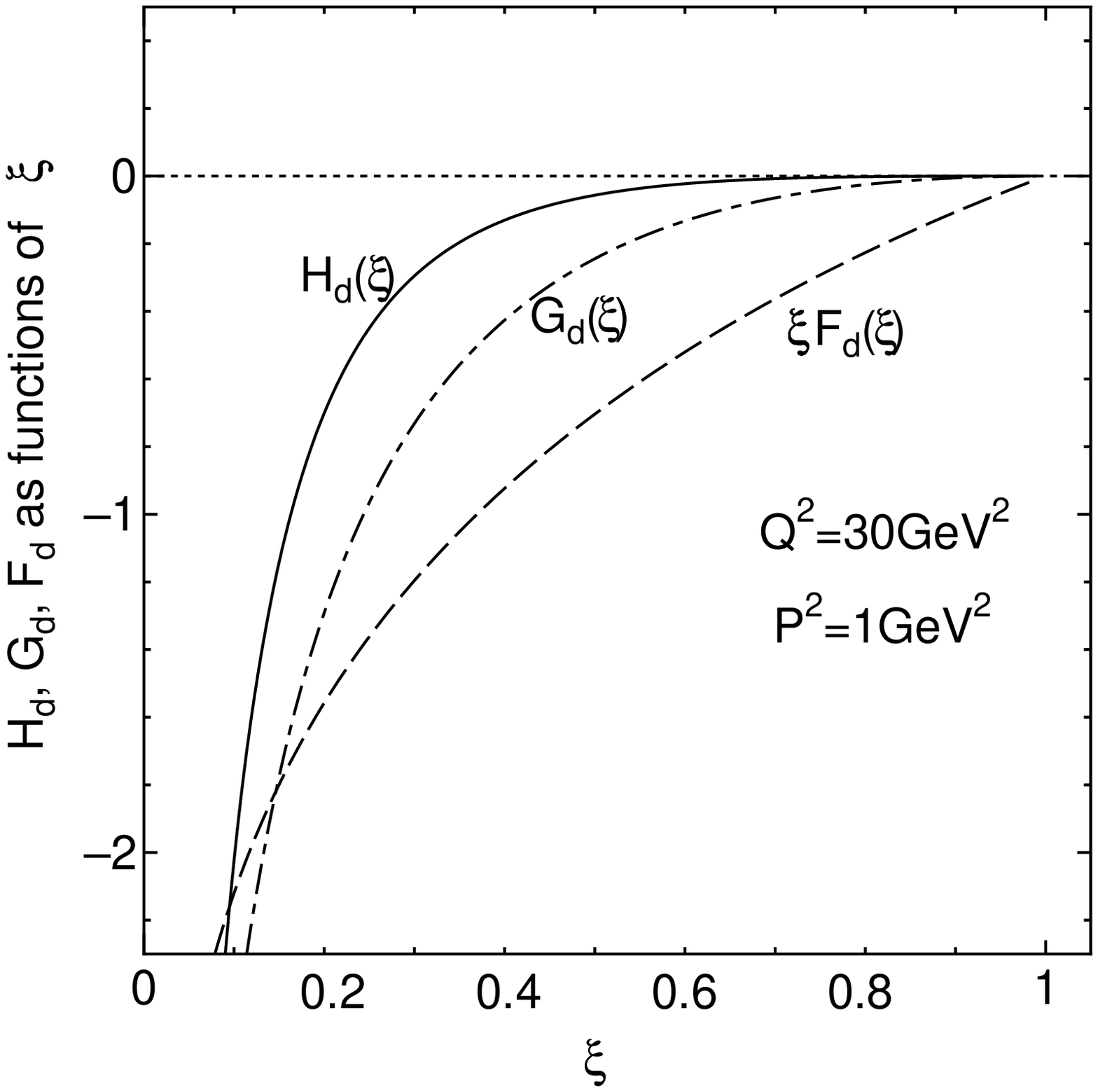}
\vspace{+3cm}
\centerline{\large\bf Fig. 4}
\end{center}
\end{figure}

\newpage
\pagestyle{empty}
\input epsf.sty
\begin{figure}
\begin{center}
\epsfxsize=16cm
\epsfbox{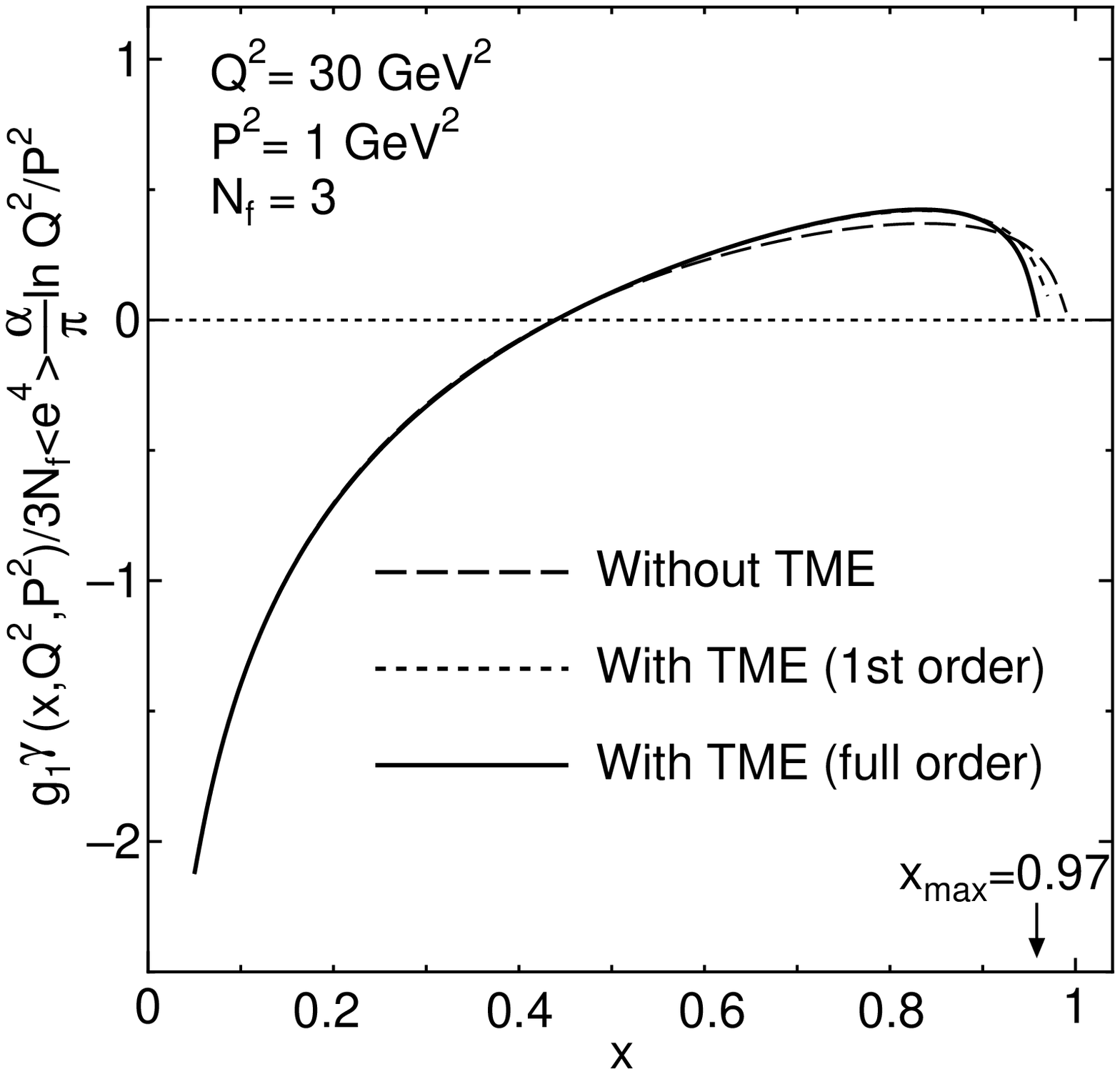}
\vspace{3cm}
\centerline{\large\bf Fig. 5}
\end{center}
\end{figure}

\newpage
\pagestyle{empty}
\input epsf.sty
\begin{figure}
\centerline{
\epsfxsize=16cm
\epsfbox{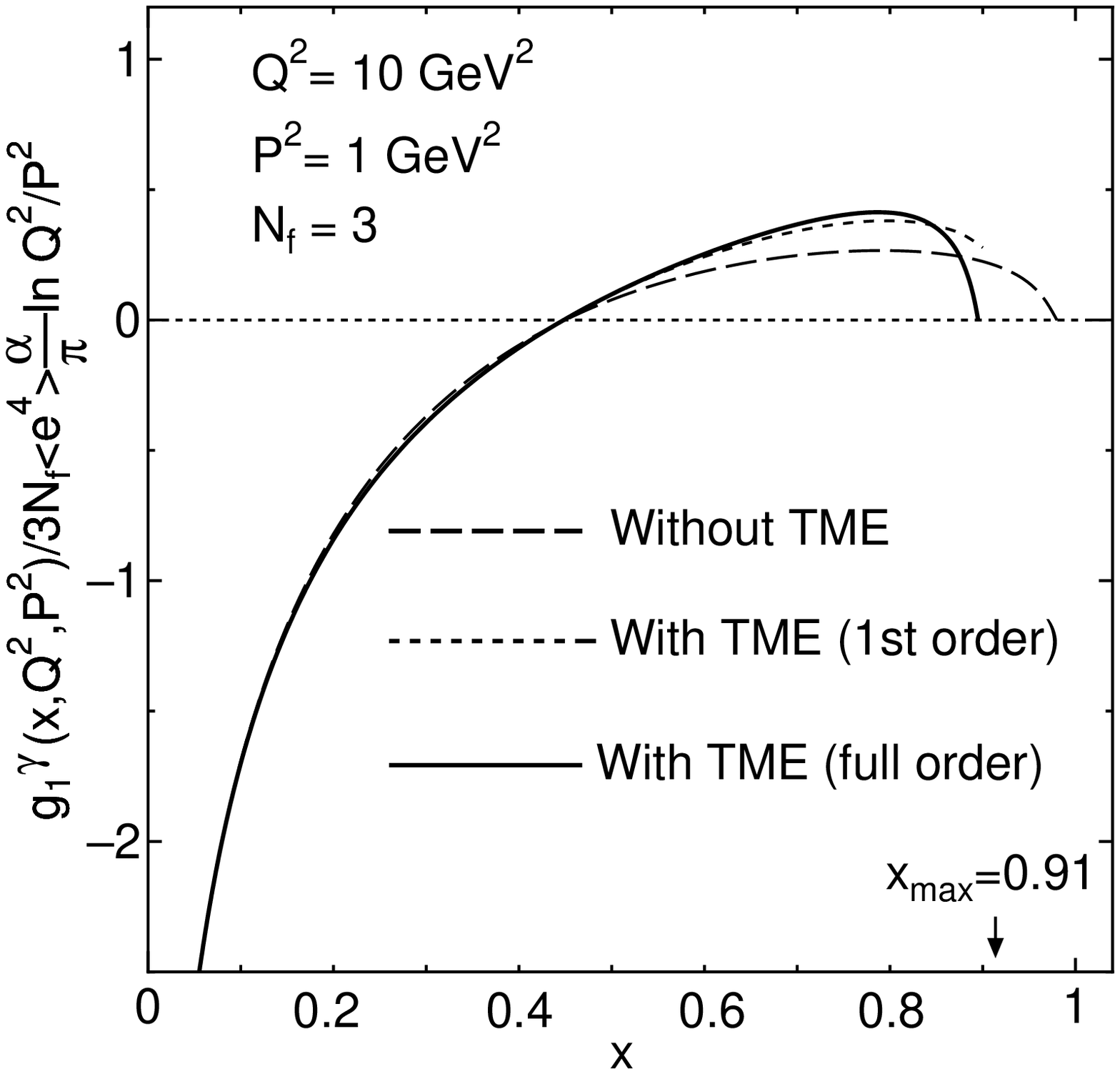}}
\vspace{+3cm}
\centerline{\large\bf Fig. 6}
\end{figure}

\newpage
\pagestyle{empty}
\input epsf.sty
\begin{figure}
\centerline{
\epsfxsize=16cm
\epsfbox{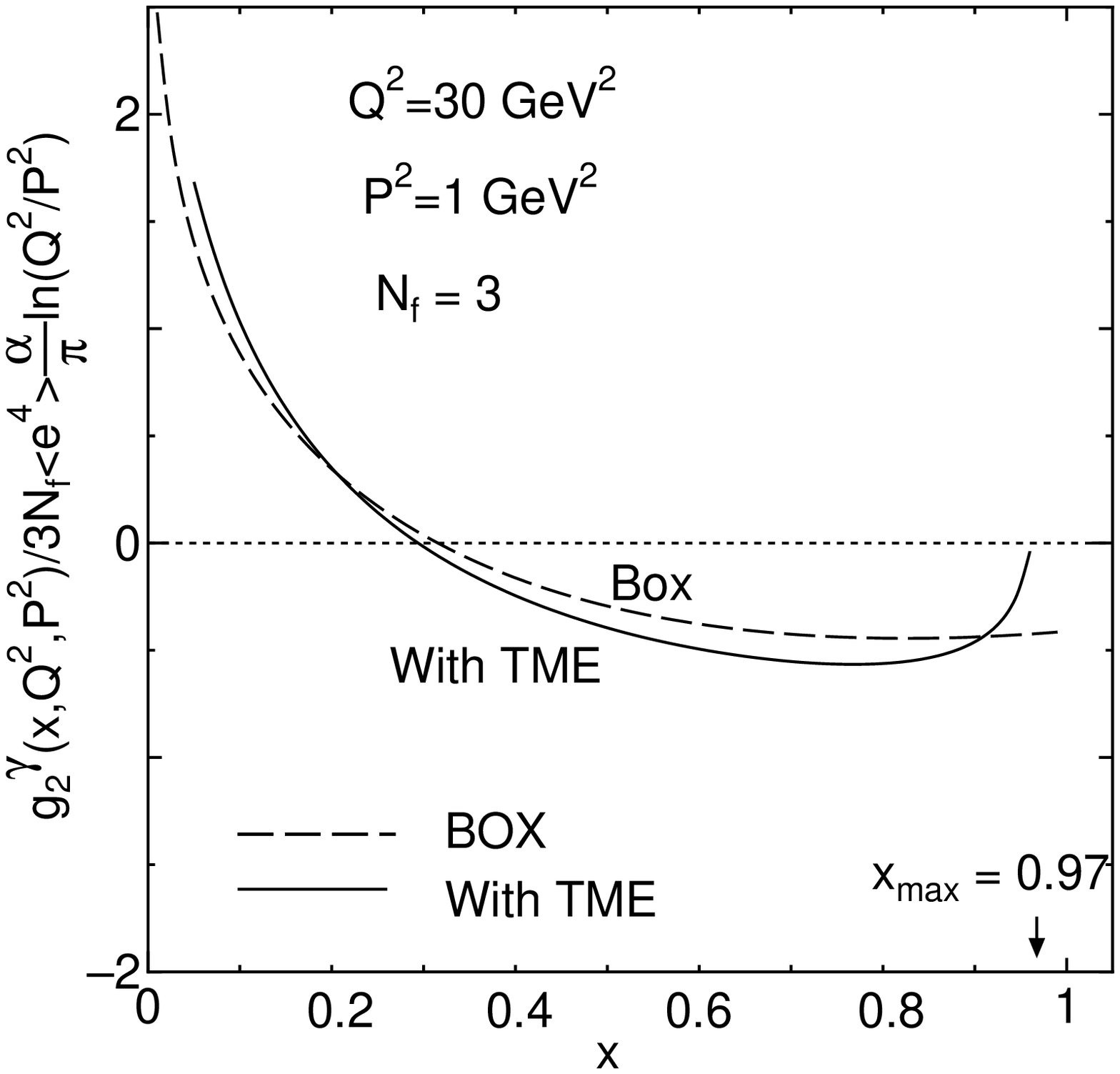}}
\vspace{+3cm}
\centerline{\large\bf Fig. 7}
\end{figure}

\newpage
\pagestyle{empty}
\input epsf.sty
\begin{figure}
\centerline{
\epsfxsize=16cm
\epsfbox{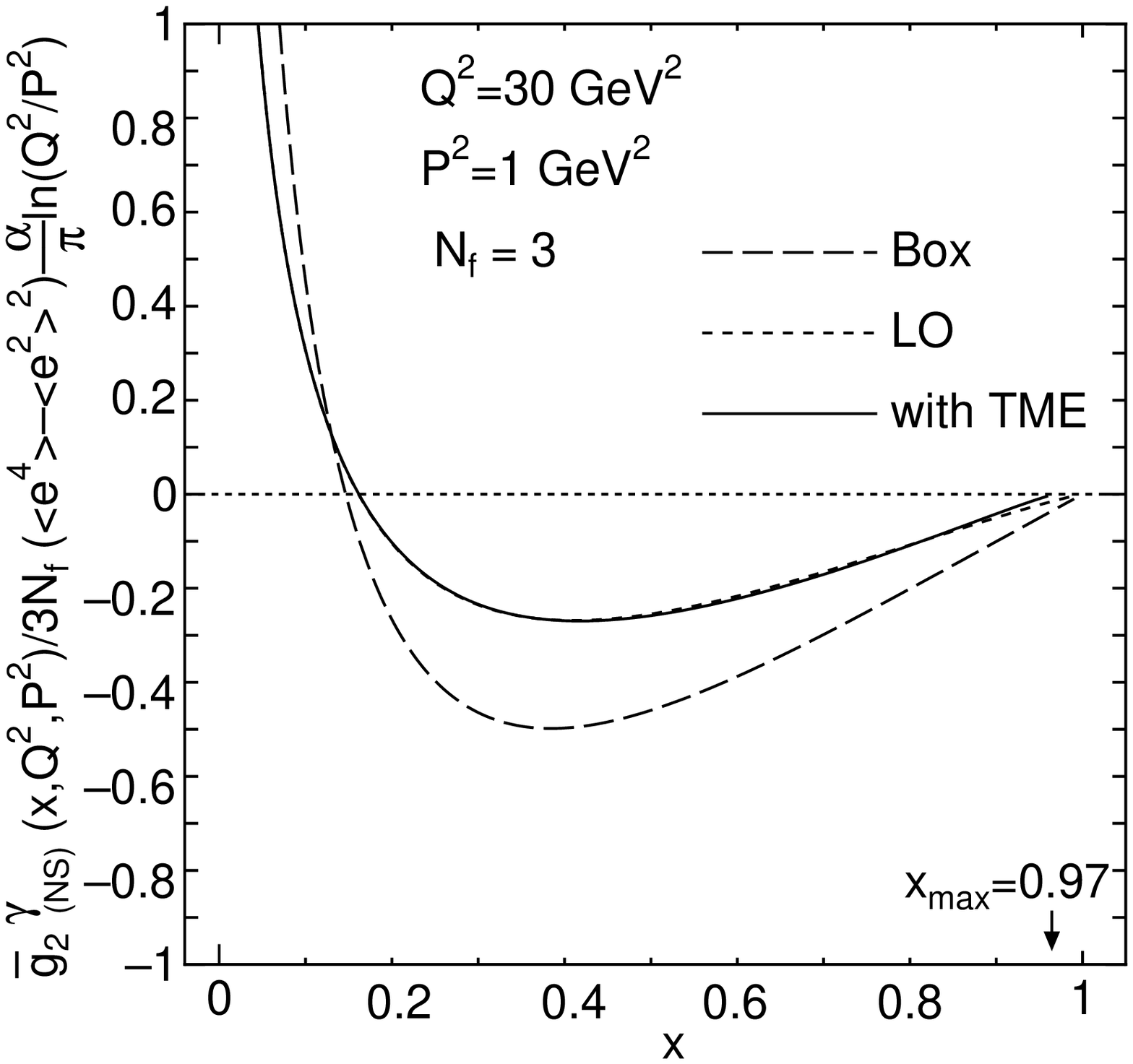}}
\vspace{+3cm}
\centerline{\large\bf Fig. 8}
\end{figure}

\newpage
\pagestyle{empty}
\input epsf.sty
\begin{figure}
\centerline{
\epsfxsize=16cm
\epsfbox{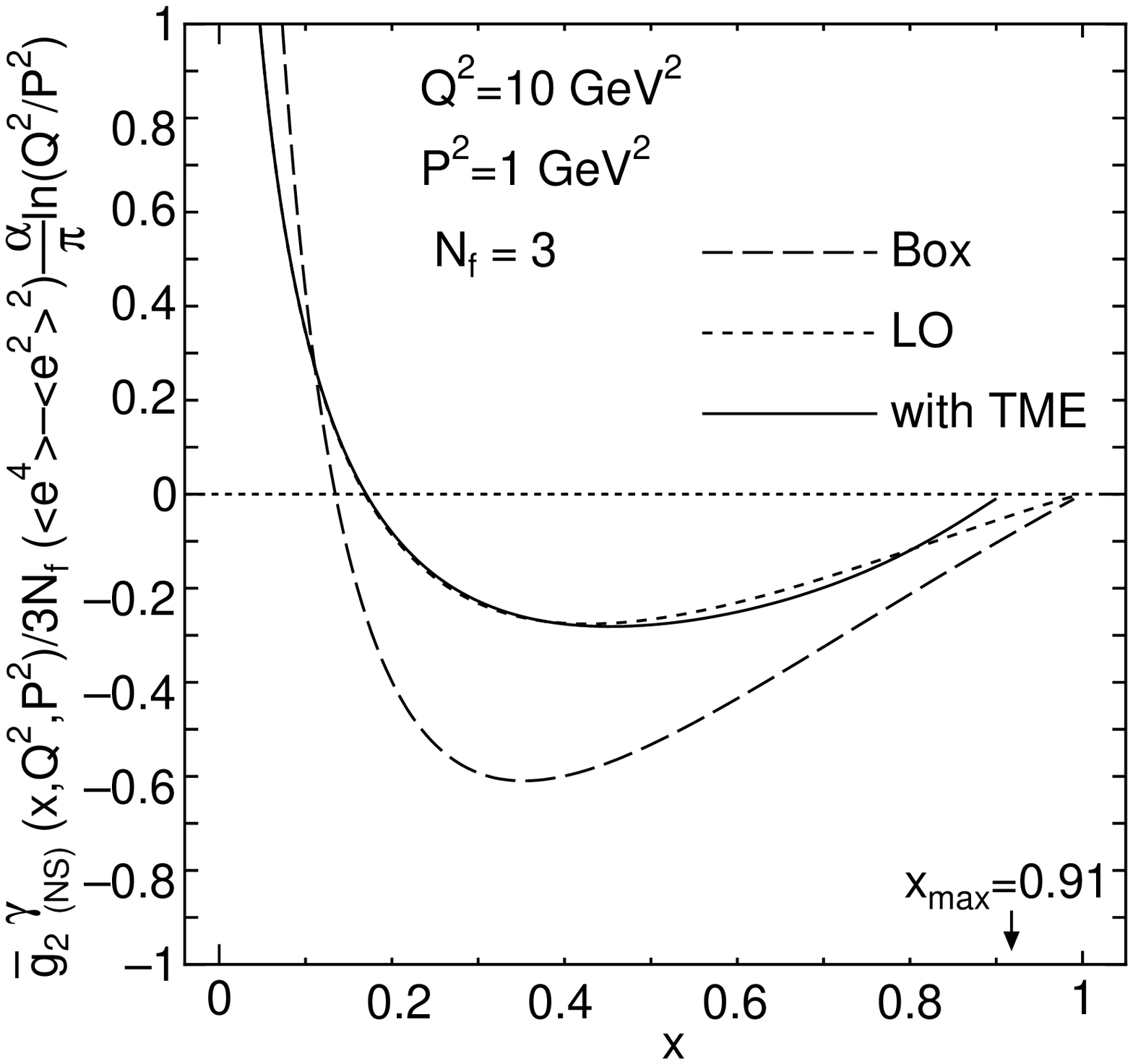}}
\vspace{+3cm}
\centerline{\large\bf Fig. 9}
\end{figure}


\end{document}